\newif\ifcheckpagelimits
\checkpagelimitstrue
\checkpagelimitsfalse

\ifcheckpagelimits
 \documentclass[nofootinbib,prl,aps,twocolumn,showpacs,showkeys,%
 amsmath,amssymb,superscriptaddress,final,reprint,floatfix,longbibliography]{revtex4-1}
 \newcommand{\todo}[1]{}
\else
 \documentclass[prl,aps,twocolumn,showpacs,showkeys,%
 amsmath,amssymb,superscriptaddress,final,reprint,floatfix,longbibliography]{revtex4-1}
 \newcommand{\todo}[1]{{\pdfmargincomment[icon=Note,color=pink]{#1}}}
\fi

\usepackage{lineno}
  \usepackage{mathptmx}
\usepackage{subfigure}
\usepackage{dcolumn}
\usepackage{amsmath,amssymb}
\usepackage{bm}
\usepackage{color}
\usepackage{overpic}
\usepackage{latexsym}
\usepackage{epstopdf}
\usepackage{xcolor}
\usepackage[english]{babel}
\usepackage{latexsym}
\usepackage{stmaryrd}

\usepackage{braket}

\definecolor{mygrey}{gray}{0.35}
\definecolor{myblue}{rgb}{0.2,0.2,0.8}
\definecolor{myzard}{cmyk}{0,0,0.05,0}
\definecolor{mywhite}{rgb}{1,1,1}
\definecolor{myred}{rgb}{1,0.,0.3}

\usepackage[colorlinks=true,citecolor=myblue,linkcolor=myred]{hyperref}
\DeclareMathAlphabet{\mathpzc}{OT1}{pzc}{m}{it}

\def\beq{\begin{equation}}
\def\eeq{\end{equation}}

\def\barray{\begin{eqnarray}}
\def\earray{\end{eqnarray}}

\usepackage{letltxmacro}
\LetLtxMacro{\ORIGselectlanguage}{\selectlanguage}
\makeatletter
\DeclareRobustCommand{\selectlanguage}[1]{%
  \@ifundefined{alias@\string#1}
    {\ORIGselectlanguage{#1}}
    {\begingroup\edef\x{\endgroup
       \noexpand\ORIGselectlanguage{\@nameuse{alias@#1}}}\x}%
}
\newcommand{\definelanguagealias}[2]{%
  \@namedef{alias@#1}{#2}%
}
\makeatother

\definelanguagealias{en}{english}

\begin{document}

\title{Operator content of entanglement spectra after global quenches 
in the transverse field Ising chain
}

\author{Jacopo Surace}
\affiliation{Department of Physics and SUPA, University of Strathclyde, Glasgow G4 0NG, United Kingdom}
\affiliation{ICFO-Institut de Ciències Fotòniques, The Barcelona Institute of Science and Technology, 08860 Castelldefels (Barcelona), Spain}
\author{Luca Tagliacozzo}
\affiliation{Department of Physics and SUPA, University of Strathclyde, Glasgow G4 0NG, United Kingdom}
\affiliation{Departament de F\'{\i}sica Qu\`antica i Astrof\'{\i}sica and Institut de Ci\`encies del Cosmos (ICCUB), Universitat de Barcelona,  Mart\'{\i} i Franqu\`es 1, 08028 Barcelona, Catalonia, Spain}
\author{Erik Tonni}
\affiliation{SISSA and INFN Sezione di Trieste, via Bonomea 265, 34136 Trieste, Italy}

\begin{abstract}
We consider the time evolution of the gaps of the entanglement spectrum 
for a block of consecutive sites in finite transverse field Ising chains 
after sudden quenches of the magnetic field. 
We provide numerical evidence that,
whenever we quench at or across the quantum critical point,
the time evolution of the ratios of these gaps allows to obtain universal information. 
They encode the low-lying gaps of the conformal spectrum 
of the Ising boundary conformal field theory 
describing the spatial bipartition 
within the imaginary time path integral approach to global quenches
at the quantum critical point.

\end{abstract}

\ifcheckpagelimits\else
\maketitle
\fi

\ifcheckpagelimits
\else

Quantum many-body systems described by local Hamiltonians
are difficult to solve because  the dimensionality of their Hilbert space increases exponentially with the number of their constituents. 
At equilibrium however the wave-functions of quantum many-body systems described by gapped Hamiltonians contain a limited amount of entanglement with a well defined structure that can  be used to perform numerical simulations with tensor networks \cite{hastings2007,masanes2009,amico2008,bridgeman2017,orus2014b, ran2017}.
 At quantum critical points (QCP), where the Hamiltonians become gapless,
the entanglement grows following universal laws \cite{pelissetto2002} that in the specific case of conformal invariant QCP
allow to unveil the data of the underlying conformal field theory (CFT)
\cite{belavin1984,henkel1999}.
In two spacetime dimensions, these data include  the central charge $c$,
the exponents dictating the decay of correlation functions that constitute the conformal spectrum 
and the coefficients of the operator product expansion.
The central charge $c$ can be read from the scaling of the
entanglement entropy (EE) of an interval with respect to its size
\cite{holzhey1994,calabrese2004,vidal2003a,callan1994b}.
The remaining CFT data are encoded e.g.
in the EE when $A$ is made by disjoint intervals
\cite{calabrese2009b,calabrese2011b,alba2011,coser2014a, denobili2015},
in a complicated way. 
The entanglement spectrum (ES) \cite{li2008,peschel1987, calabrese2008,poilblanc2010,cirac2011,dubail2012,lauchli2013a},
i.e. the spectrum of the reduced density matrix,
can be related to the conformal spectrum of a 
conformal field theory with boundaries (BCFT)
defined by proper conformal boundary conditions (CBC)
\cite{cardy2016a},
as also observed by L\"auchli in numerical studies
\cite{lauchli2013a}.

Entanglement plays a central role in our understanding of large many-body quantum systems, as witnessed by the numerous proposals on how to measure entanglement in experiments which have led to the first experimental measures in the context of cold atoms and trapped ions both at and out of equilibrium 
\cite{cardy2011,abanin2012,daley2012,islam2015,hauke2016,pichler2016,kaufman2016,brydges2019,lukin2019}.  
Such experiments are expected to play an increasingly important role  out of equilibrium, where the amount of entanglement increases to its maximum value very fast with time \cite{calabrese2005c,dechiara2006,nahum2017}. This creates an entanglement barrier that prevents to extract reliable predictions at long time  even when using our best  classical-simulation techniques
\cite{vidal2004,white2004,daley2004,paeckel2019a} 
(see \cite{leviatan2017a,white2017,surace2019,rams2019,krumnow2019} 
for recent proposals to overcome this barrier).

As a result, important questions about strongly correlated quantum many-body quantum systems out of equilibrium
are only partially understood and currently can only be settled experimentally.
Examples of open questions include
the existence of new phases of matter out of equilibrium \cite{fausti2011,abanin2017,alet2018a}, the presence/absence of equilibration, 
the phenomenon of thermalization or generalized thermalisation that should reconcile the reversibility of quantum mechanics 
with the irreversibility of the macroscopic world \cite{rigol2008,polkovnikov2011,eisert2015, dalessio2016}. A possible hope is that also out of equilibrium a form of universality exists. This would allow for example to obtain at least quantitative information even with approximate algorithms.

Out of equilibrium universality has been observed only in some specific scenarios where the scaling of the spatio-temporal evolution of the system 
can be described by  universal exponents and functions (see the recent experiment  \cite{prufer2018a} and the references therein). 
Dynamical quantum phase transitions, appearing as singularities in the return probabilities to the initial state quenched across a quantum critical point, 
could also provide a scaling region that could be explained through out-of-equilibrium universality \cite{heyl2013, zunkovic2018, heyl2019}.

In this Letter we consider the
spin-$\tfrac{1}{2}$ transverse field Ising chain (TFIC),
a paradigmatic model for one-dimensional quantum critical behaviour 
and quantum phase transitions.
We study the time evolutions of the ES 
after a global quench 
given by a sudden change of the magnetic field,
when the subsystem $A$  is an interval in a finite system.
We provide numerical evidence that,
for quenches to or across the QCP,
the gaps of the ES carry information about the 
conformal spectrum of the Ising BCFT
describing the QCP.
These results are quite robust 
under changes of the quench protocol thus providing a new example of universality out of equilibrium.


\paragraph{Setup.}
The Hamiltonian of the TFIC is
\begin{equation}
\label{ising-chain-hamiltonian}
H(\theta) =
- \,\frac{1}{2} \left(\, \sum_{i=1}^{L-1} \sigma_i^x \sigma_{i+1}^x  + \cot\theta \sum_{i=1}^{L} \sigma_i^z +\eta \, \sigma^x_L \sigma^x_1 \right)
\end{equation}
where $\sigma_i^\alpha$ are the Pauli matrices at the $i$-th site 
and $\cot\theta$ is the magnetic field, with $0< \theta <\tfrac{\pi}{2}$.
At zero temperature and in the thermodynamic limit, this model exhibits 
a ferromagnetic (ordered) phase for $\tfrac{\pi}{4} < \theta < \tfrac{\pi}{2} $ 
and a paramagnetic (disordered) phase for $0 < \theta< \tfrac{\pi}{4} $,
separated by a QCP  at $\theta =\tfrac{\pi}{4}$.
We consider finite systems made by $L$ sites 
where either periodic boundary conditions (PBC)
or free open boundary conditions (OBC) are imposed,
which correspond respectively to $\eta=1$ and $\eta=0$ 
in (\ref{ising-chain-hamiltonian}).

Entanglement can be studied by partitioning the system into an interval $A$ 
made by $\ell < L/2$ consecutive sites and its complement $B$
(in the case of OBC, $d$ denotes the number of sites 
on the left of $A$).
Assuming that the Hilbert space can accordingly be written  as
$\mathcal{H} = \mathcal{H}_A \otimes \mathcal{H}_B$,
we are interested in the ES,
i.e. in the eigenvalues $\lambda_i \equiv e^{-\xi_i} $ (where $\xi_i \geqslant 0$)
of the reduced density matrix $\rho_A = \textrm{Tr}_{\mathcal{H}_B} \rho$
(normalised by $\textrm{Tr}_{\mathcal{H}_A} \rho_A = 1$) of $A$.
In order to get rid of arbitrary entanglement energy shifts we consider the gaps
$g_{r} \equiv 
\log \lambda_{\textrm{\tiny max}} - \log \lambda_r 
= \xi_r - \xi_{\textrm{\tiny min}} \geqslant 0 $ 
with $r\geqslant 1$.
The EE  of $A$ is
$S_A =  - \textrm{Tr}_{\mathcal{H}_A}( \rho_A \log   \rho_A)$. 
Given the system in the ground state $|\psi_0 \rangle$ 
of (\ref{ising-chain-hamiltonian}) for $\theta = \theta_0$,
at $t=0$ the magnetic field is changed abruptly $\theta_0 \to \theta$,
hence the time evolved state is 
$| \psi(t) \rangle=e^{- \textrm{i}\, H( \theta) \,t} | \psi_0 \rangle $.

In spectroscopy the ratios $g_r/g_s$ are usually studied
in order to get rid of a shift and a rescaling of the entire spectrum.
These quantities are suggested also by the analytic expressions 
for the time evolution of the ES of half of an infinite line obtained through CFT methods when the post-quench Hamiltonian is critical
\cite{cardy2016a},
which tell us that $(\ell g_r)^{-1}$ and $g_r S_A$  depend on  critical exponents.
We study the time evolutions of $g_r/g_1$, of $(\ell g_r)^{-1}$
and of $g_r S_A$ after global quenches of the magnetic field  in the TFIC
for various $\theta_0$ and $\theta$,
showing the numerical results for the first $16$ eigenvalues
(hence $r=1,2, \dots, 15$). 
The numerical analysis is performed by using the standard mapping 
of the TFIC to a chain of free fermions 
\cite{jordan1928,lieb1961,barouch1970,barouch1971,barouch1971a}
(see also \cite{surace2019} and references therein). 
The time evolutions of these quantities after some global quenches 
have been recently studied in an infinite harmonic chain 
or in an infinite chain of free fermions at half filling
\cite{digiulio2019}.

 \begin{figure}[t!]
\vspace{-.0cm}
\begin{center}
\includegraphics[width=1.\columnwidth]{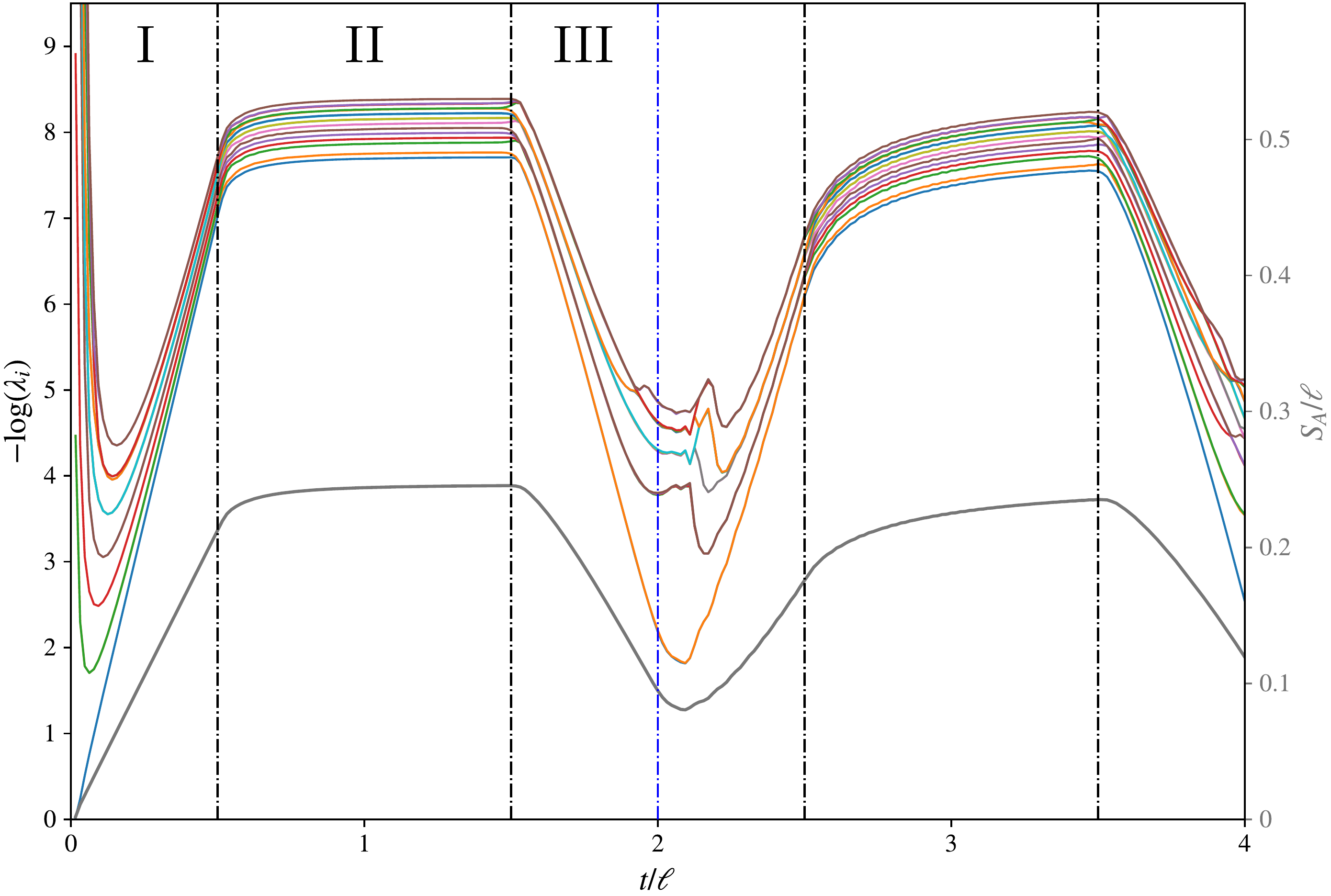}
\end{center}
\vspace{-.6cm}
\caption{(color online).
Time evolutions of the first 16 eigenvalues 
$\lambda_{\textrm{\tiny max}} \geqslant \lambda_1 \geqslant \lambda_2 \geqslant  \dots \geqslant \lambda_{15}$
of the ES and of the EE (grey line)
for an interval with $\ell =64$ sites in the chain with $L=256$ sites and PBC
after the quench $\theta = \pi/8 \to \theta = \pi/4$ to the QCP.
Different degeneracies are observed in the regimes I, II and III
within the period.
}
\label{fig:ES-typical}
\end{figure}


\paragraph{Quenches at the QCP.}
We report numerical results for
the quench $\theta_0 = \pi/8 \to \theta = \pi/4$,
from the ferromagnetic phase to the quantum critical point, since they are paradigmatic for generic quenches to the critical point obtained for different choices of $\theta_0$.
The interval $A$ is
either in the chain with PBC 
(Fig.\,\ref{fig:ES-typical} and Fig.\,\ref{fig:ToCriticalPBC})
or in the chain with OBC and $d=0$ (Fig.\,\ref{fig:ToCriticalOBC})
or 
$d=\tfrac{3}{2}\ell$ 
(Fig.\,\ref{fig:ToCriticalOBCnotsymm}).
In the time evolutions of the ES and of the EE in Fig.\,\ref{fig:ES-typical},
we see several time regimes 
(separated by black dashed-dotted vertical lines),
that can be identified by employing 
the quasi-particle picture \cite{calabrese2005c,calabrese2006a,calabrese2007}.
It assumes that
each spatial point emits a pair of oppositely moving 
and initially entangled quasi-particles
that propagate semiclassically 
with velocity $v_{\textrm{\tiny q}} =  \textrm{min}(1, \cot\theta)$ 
after the quench \cite{robinson1976,bratteli1981}. 
In finite systems this implies quantum recurrences:
the quasi-particles emitted at the same point
meet periodically at times that are 
multiple integers of $\tfrac{L/2}{v_{\textrm{\tiny q}}}$ for PBC 
and of $\tfrac{L}{v_{\textrm{\tiny q}}}$ for OBC 
(in the figures the end of a period corresponds 
to the blue dashed-dotted vertical  line)
\cite{cardy2014,cardy2016g,takayanagi2010,dasilva2015a}.
When $A$ is an interval in the chain with PBC
(Fig.\,\ref{fig:ES-typical}, Fig.\,\ref{fig:ToCriticalPBC},
Fig.\,\ref{fig:AcrossPBC} and Fig.\,\ref{fig:InsidePhasePBC}),
we can identify three regimes  I, II, III ending respectively at
$t_1 = \tfrac{\ell / 2}{v_{\textrm{q}}}$ (equilibration time), 
$t_2= \tfrac{(L-\ell)/2}{v_{\textrm{q}}}$
and $t_3= \tfrac{L/2}{v_{\textrm{q}}}$.
When $A$ is in the chain with OBC
and $d=0$ (Fig.\,\ref{fig:ToCriticalOBC})
these values must be just multiplied by a factor of $2$,
while for $d\neq 0$ with $d+\ell < L/2$ (Fig.\,\ref{fig:ToCriticalOBCnotsymm})
the bounces of the quasi-particles at the boundaries
provide nine regimes I, II, \dots, IX within a period.

The time evolutions in Fig.\,\ref{fig:ToCriticalPBC},
Fig.\,\ref{fig:ToCriticalOBC} and Fig.\,\ref{fig:ToCriticalOBCnotsymm}
display some interesting common features. 
For $(\ell g_r)^{-1}$ (top panels)
we observe either linear growths  or plateaux or linear decreases 
in the different regimes,
in correspondence of the same behaviour in the EE,
with the crucial difference that
$(\ell g_r)^{-1}$ display a higher degree of flatness than EE
in regime II, up to oscillations.
Understanding the origin of this difference 
in terms of a Generalised Gibbs Ensemble
\cite{rigol2006, rigol2007b} (see \cite{vidmar2016} for a recent review)
could be very insightful.

As for the products $g_r S_A$ (bottom panels),
their time evolutions in regimes I and II approximatively take constant values
while they become very complicated in the regimes different from I
whenever $(\ell g_r)^{-1}$  either increase or decrease linearly.

The most interesting behaviour is displayed by the ratios $g_r/g_1$:
up to oscillations, 
they take approximatively constant values in the different time regimes
and the same constant values are observed also in the corresponding regimes of the second period. 
It is crucial to remark that these constant values are positive integers. 
Focussing on the regimes I and II, we stress that 
the integer $1$ is observed only in the regime I in the case of PBC.
Furthermore, we find it worth observing the changes of degeneracies in passing
from a regime to the subsequent one for PBC in Fig.\,\ref{fig:ToCriticalPBC}
and for OBC in Fig.\,\ref{fig:ToCriticalOBCnotsymm}
(from I to II and from VIII to IX).
In particular, for PBC, the higher degeneracy in regime III leads to observe that 
this regime looks like the time reversal of regime I of the second period. 
Similarly, the degeneracies seem to tell that  
the regime I in the first period could be related to the regime III in the second period,
maybe because for PBC the quasi-particles 
come back to their emission point at the end of the second period. 
Comparing the regimes II in Fig.\,\ref{fig:ToCriticalPBC} and Fig.\,\ref{fig:ToCriticalOBC},
we observe that the plateaux for $(\ell g_r)^{-1}$ have the same heights. 
In Fig.\,\ref{fig:ToCriticalOBCnotsymm}, 
where OBC are imposed and $d> \tfrac{\ell}{2}$,
the ratios $g_r/g_1$ take almost constant integer values and 
the regimes I and II are like in Fig.\,\ref{fig:ToCriticalPBC},
where PBC are imposed.

A remarkable feature of the time evolutions of $g_r/g_1$
in Fig.\,\ref{fig:ToCriticalPBC} and Fig.\,\ref{fig:ToCriticalOBC}
is their very mild dependence on the initial state within the paramagnetic phase:
changing $0 < \theta_0 < \pi /2$
(we have considered $\theta_0 = \pi/13$ and $\theta_0 = \pi/10$ \cite{j.suracea})
the same integer constant values with the same degeneracies occur.
For PBC the same observation can be made
taking initial states within the ferromagnetic phase
(we have considered $\theta_0 = \pi/2-\pi/13$ and $\theta_0 = \pi/2- \pi/10$):
only the degeneracies change with respect to the cases with $0 < \theta_0 < \pi /2$.
Instead, for OBC when $d=0$, taking different $ \pi/4 < \theta_0 < \pi/2$
we find that the regime II is the same, while in the regime I 
all the integer values (including $1$) are taken by $g_r/g_1$,
and that the degeneracies change passing from regime I to regime II.
Thus, regime II in the time evolutions of the ratios $g_r/g_1$ in 
Fig.\,\ref{fig:ToCriticalPBC} and Fig.\,\ref{fig:ToCriticalOBC}
is independent of the initial state. A mild dependence on the initial state occurs in the peak that $g_r/g_1$ display at the beginning of regime 1. 
Furthermore, $g_r S_A$ take different 
approximatively constant values for different initial states, 
as observed also in harmonic chains \cite{digiulio2019}.

\begin{figure}[t!]
\vspace{-.0cm}
\begin{center}
\includegraphics[width=1.05\columnwidth]{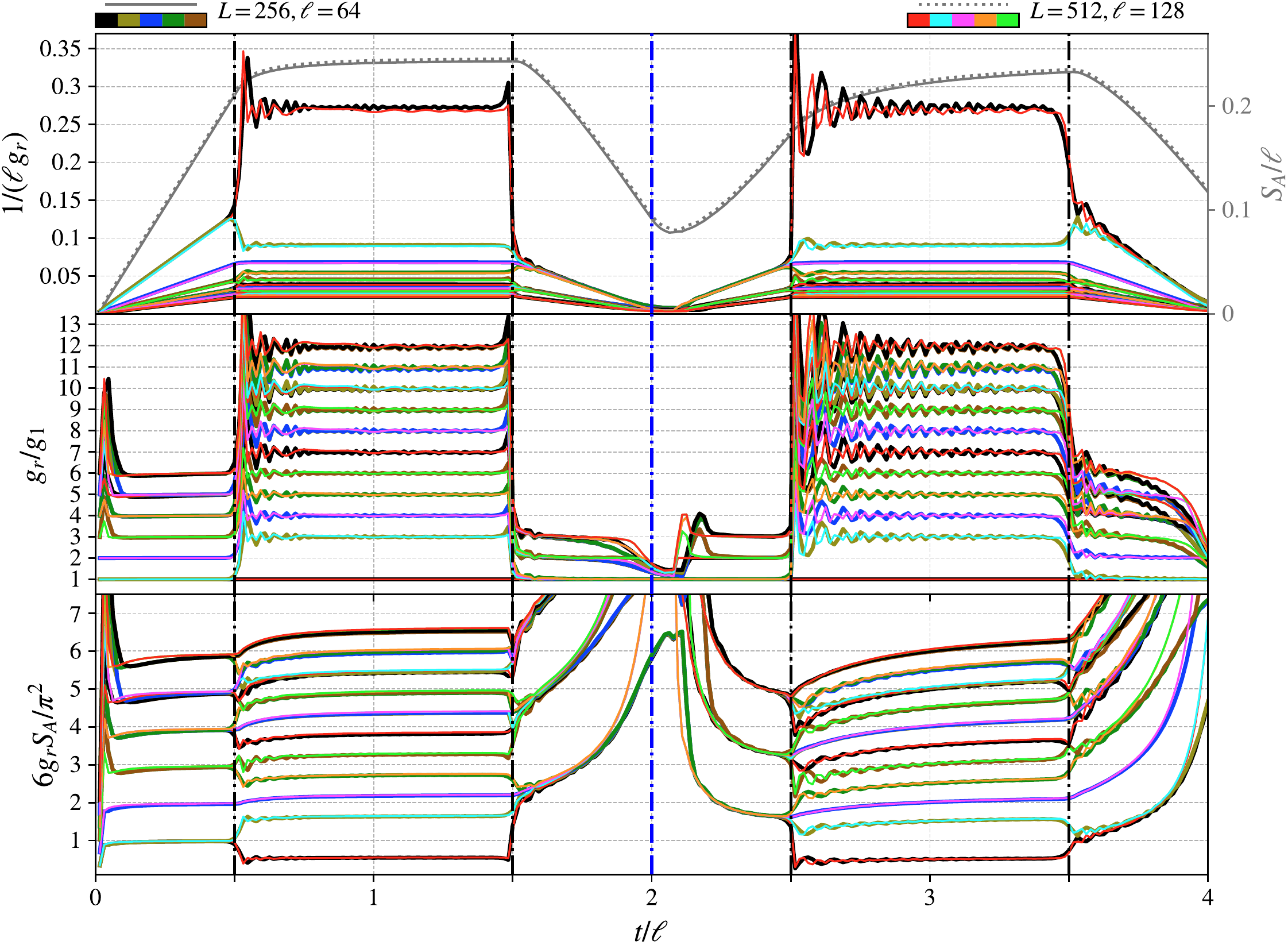}
\end{center}
\vspace{-.6cm}
\caption{(color online).
Time evolution of $(\ell g_r)^{-1}$ (top), $g_r/g_1$ (middle) and $g_r S_A$ (bottom)
after the quench $\theta_0 = \pi/8 \to \theta = \pi/4$
for an interval in the chain with PBC.
}
\label{fig:ToCriticalPBC}
\end{figure}

\paragraph{Insights from BCFT.}
Analytic CFT expressions for the time evolutions of the ES
in Fig.\,\ref{fig:ToCriticalPBC}, Fig.\,\ref{fig:ToCriticalOBC} and Fig.\,\ref{fig:ToCriticalOBCnotsymm}
are not available in the literature.
Nonetheless, some of their characteristic features
can be justified through the CFT analysis performed in \cite{cardy2016a}
for the ES of half of the infinite line,
based on BCFT \cite{cardy1986a,cardy1989}.
The continuum limit of the TFIC at the QCP 
is described by the Ising CFT, whose central charge is $c= \tfrac{1}{2}$.
In the presence of boundaries, 
the conformal symmetry of the Ising BCFT allows
only three CBC.

In the BCFT approach to global quantum quenches 
with critical evolution Hamiltonians
\cite{calabrese2005c,calabrese2006a,calabrese2007,cardy2014,cardy2016g}
(see \cite{calabrese2016a} for a recent review),
the spacetime to consider in the imaginary time path integral
is a strip whose width is given by $\tau_0$ along the imaginary time direction 
and by the size of the system along the spatial direction.
On both the lines delimiting the strip in the imaginary time direction,
the state $| \psi_0 \rangle$ is approximated 
by the conformal boundary state $| \tilde{\psi}_0 \rangle$
corresponding to some CBC $b_0$,
while the CBC $b_1$ and $b_2$ are imposed 
along the parallel segments 
corresponding to the physical boundaries
(for the OBC that we consider $b_1 = b_2 \equiv b$,
given by free boundary conditions).
 \begin{figure}[t!]
\vspace{-.0cm}
\begin{center}
\includegraphics[width=1.05\columnwidth]{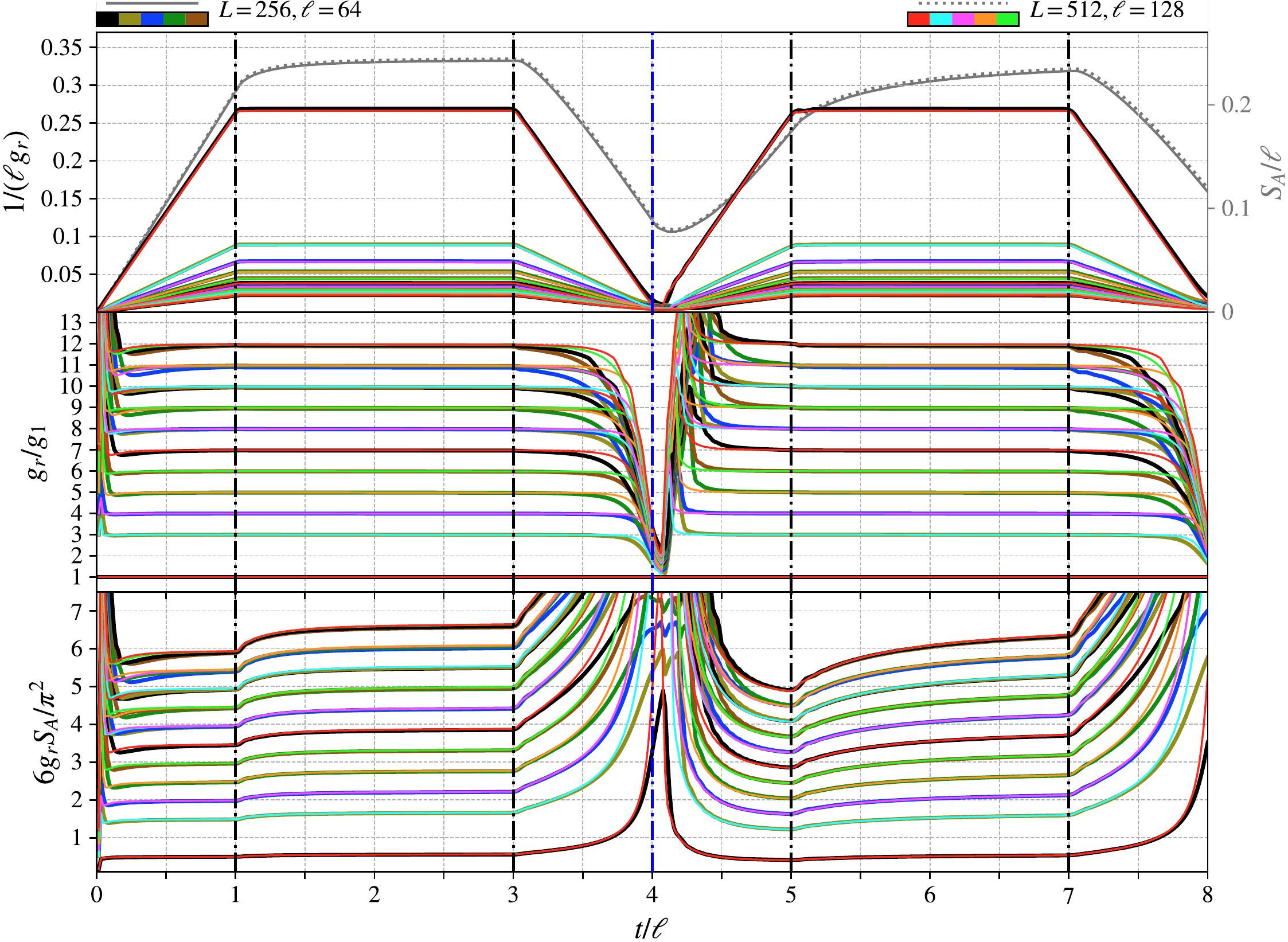}
\end{center}
\vspace{-.6cm}
\caption{(color online).
Time evolution of $(\ell g_r)^{-1}$ (top), $g_r/g_1$ (middle) and $g_r S_A$ (bottom)
after the quench $\theta_0 = \pi/8 \to \theta = \pi/4$
for an interval 
in the chain with OBC, for $d=0$.
}
\label{fig:ToCriticalOBC}
\end{figure}
 \begin{figure}[b!]
\vspace{-.3cm}
\begin{center}
\includegraphics[width=1.05\columnwidth]{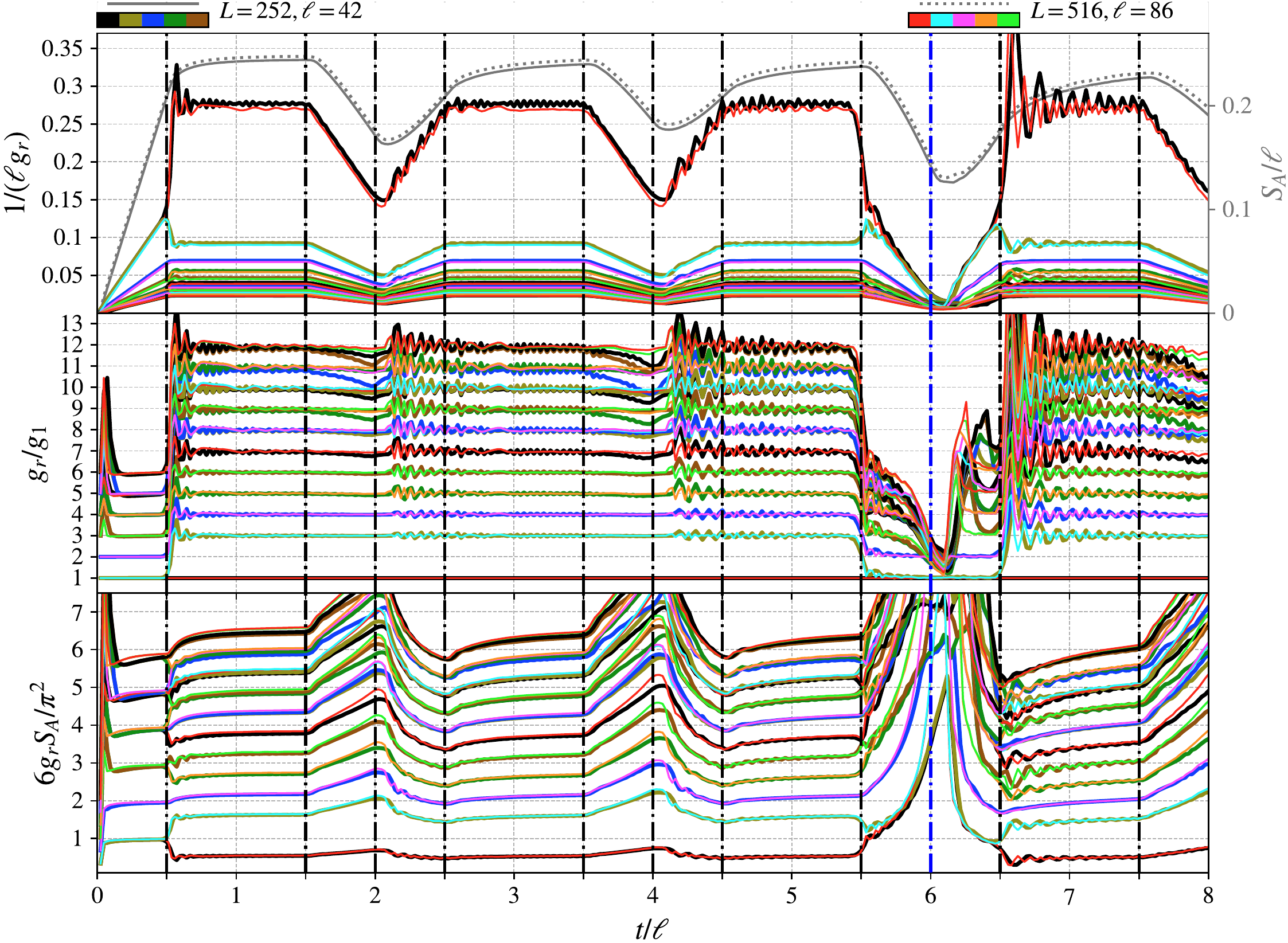}
\end{center}
\vspace{-.6cm}
\caption{(color online).
Time evolution of $(\ell g_r)^{-1}$ (top), $g_r/g_1$ (middle) and $g_r S_A$ (bottom)
after the quench $\theta_0 = \pi/8 \to \theta = \pi/4$
for an interval made by $\ell$ sites in the chain with OBC, for $d=\tfrac{3}{2}\ell$.
}
\label{fig:ToCriticalOBCnotsymm}
\end{figure}
Following \cite{cardy2016a}
(see also \cite{holzhey1994,peschel2004,ohmori2015a,alba2017b,tonni2018,wen2018a}), 
it is convenient to regularise 
the ultraviolet (UV) divergencies 
by removing small disks centered at the entangling points
whose radius is the infinitesimal UV cutoff $\epsilon$.
This introduces a boundary around each entangling point, 
where the CBC $a$ is imposed.

When the subsystem $A$ is half of the infinite line,
the spacetime can be conformally mapped into an annulus 
whose CBC are $a$ and $b_0$.
After a crucial analytic continuation to real time,
in this case one finds
$1/g_r = (2\pi \tau_0 \,\Delta_r)^{-1} t $ grow linearly in time 
and $g_r/g_1 = \Delta_r / \Delta_1$ are independent of time, 
where $r\geqslant 1$ and $\Delta_r \in \mathcal{S}(a, b_0)\setminus \{0\}$,
being $ \mathcal{S}(a, b_0)$ the conformal spectrum
(made by the conformal dimensions of the primary fields and of their descendants)
of the BCFT on the annulus with CBC given by $a$ and $b_0$ 
\cite{cardy2016a}.
The parameter $\tau_0$ encodes information about the initial state. 
Furthermore, the linear growth of the EE for this bipartition
\cite{calabrese2005c,calabrese2006a,calabrese2007}
leads to  $g_r \,S_A  = \tfrac{\pi^2c}{3}\, \Delta_r$, which is independent of time. 
We remark that $g_r/g_1$ and $g_r \,S_A $ are independent of $\tau_0$.

When the system is finite and the subsystem $A$ has 
$N$ entangling points ($N$ must be even for PBC),
the spacetime has the topology of a sphere 
with $N + 2$ boundaries for PBC
and with $N + 1$ boundaries for OBC.
We expect that, at the beginning (regime I),
the time evolution of the ES is determined 
only by the spacetime around the entangling points. 
This assumption leads to consider $N$ independent copies 
of the spacetime corresponding to the time evolution of
half of the infinite line;
hence to employ the conformal spectrum $\mathcal{S}_N(a,b_0)$ 
of $N$ independent copies of the same BCFT on the annulus 
with CBC given by $a$ and $b_0$.
Combining this assumption with the results of \cite{cardy1989}
and assigning free boundary conditions to both $a$ and $b_0$,
we can explain the data observed in the regimes I
of Fig.\,\ref{fig:ToCriticalPBC}, Fig.\,\ref{fig:ToCriticalOBC} and Fig.\,\ref{fig:ToCriticalOBCnotsymm},
where $N=2$, $N=1$ and $N=2$ respectively.
This assignment for $a$ has been found also 
through the numerical analysis of the ES
for some bipartitions at equilibrium \cite{lauchli2013a}.
We remark that also $b$ corresponds to free boundary conditions in our numerical analysis.
These assignments lead to
$\mathcal{S}(a, b_0) = \mathcal{S}(a, a) = \mathcal{S}(a, b) = 
\{ 0, \tfrac{1}{2}, \tfrac{3}{2},2, \tfrac{5}{2}, 3, \tfrac{7}{2}, 4, \tfrac{9}{2}, 5, \dots   \}$
(for recent numerical results see \cite{evenbly2010b,evenbly2014}),
which implies that
$\mathcal{S}_N(a, b_0)= 
\{ 0, \tfrac{1}{2}, 1, \tfrac{3}{2},2, \tfrac{5}{2}, 3, \tfrac{7}{2}, 4, \tfrac{9}{2}, 5,\dots   \}$ for any $N$.
For the regime I, this naive BCFT analysis leads to
$g_r \,S_A  = \tfrac{\pi^2c_{\textrm{\tiny tot}}}{3}\, \Delta_r$ 
where $c_{\textrm{\tiny tot}} = Nc$ 
and $ \Delta_r \in \mathcal{S}_N(a,b_0) \setminus\{0\}$.
The independence of time is observed from the data, 
but the numerical values of the constants depend on the initial state, as remarked above. 
This dependence on the initial state could be justified by arguing that 
$S_A$ and the gaps $g_r$ lead to different numerical values for $\tau_0$,
as already noticed for the linear growths of the R\'enyi entropies 
with different values of the R\'enyi index \cite{coser2014b}.

The quasi-particle picture allows to argue that
the regimes II are not influenced 
by the finite size of the system
for the chosen bipartitions, where $\ell < L/2$.
It also identifies the boundaries of the finite 
Euclidean spacetime that presumably influence
the time evolutions in this time regime.
For all the chosen bipartitions we just need 
the conformal spectrum of a single Ising BCFT on the annulus. 
In particular, for Fig.\,\ref{fig:ToCriticalPBC} 
and Fig.\,\ref{fig:ToCriticalOBCnotsymm} (where $d> \ell /2$)
$\mathcal{S}(a,a)$ has to be considered
because the relevant boundaries are the 
infinitesimal circles around the entangling points.
Instead, in Fig.\,\ref{fig:ToCriticalOBC} the physical 
boundary becomes important in regime II and 
$\mathcal{S}(a,b)$ has to be employed
(see also \cite{wen2018a} for this case).
In our numerical analysis both $a$ and $b$
correspond to free boundary conditions,
hence they are not distinguishable.
These observations lead to expect that
all the regimes II are identical and 
this is confirmed in Fig.\,\ref{fig:ToCriticalPBC}, 
Fig.\,\ref{fig:ToCriticalOBC} 
and Fig.\,\ref{fig:ToCriticalOBCnotsymm},
up to oscillations.
Summarising, the time evolutions of $g_r/g_1$ 
when $0<\theta_0 <\pi/4$, for the bipartitions considered,
seem determined
by the gaps of the conformal spectrum given by 
$(0 \oplus \tfrac{1}{2})^N$ in the regime I 
and by $0 \oplus \tfrac{1}{2}$ in the regime II. 
The above considerations based on BCFT are expected to hold
only for the low-lying part of the ES \cite{a.polkovnikov}.

 \begin{figure}[t!]
\vspace{-.0cm}
\begin{center}
\includegraphics[width=1.04\columnwidth]{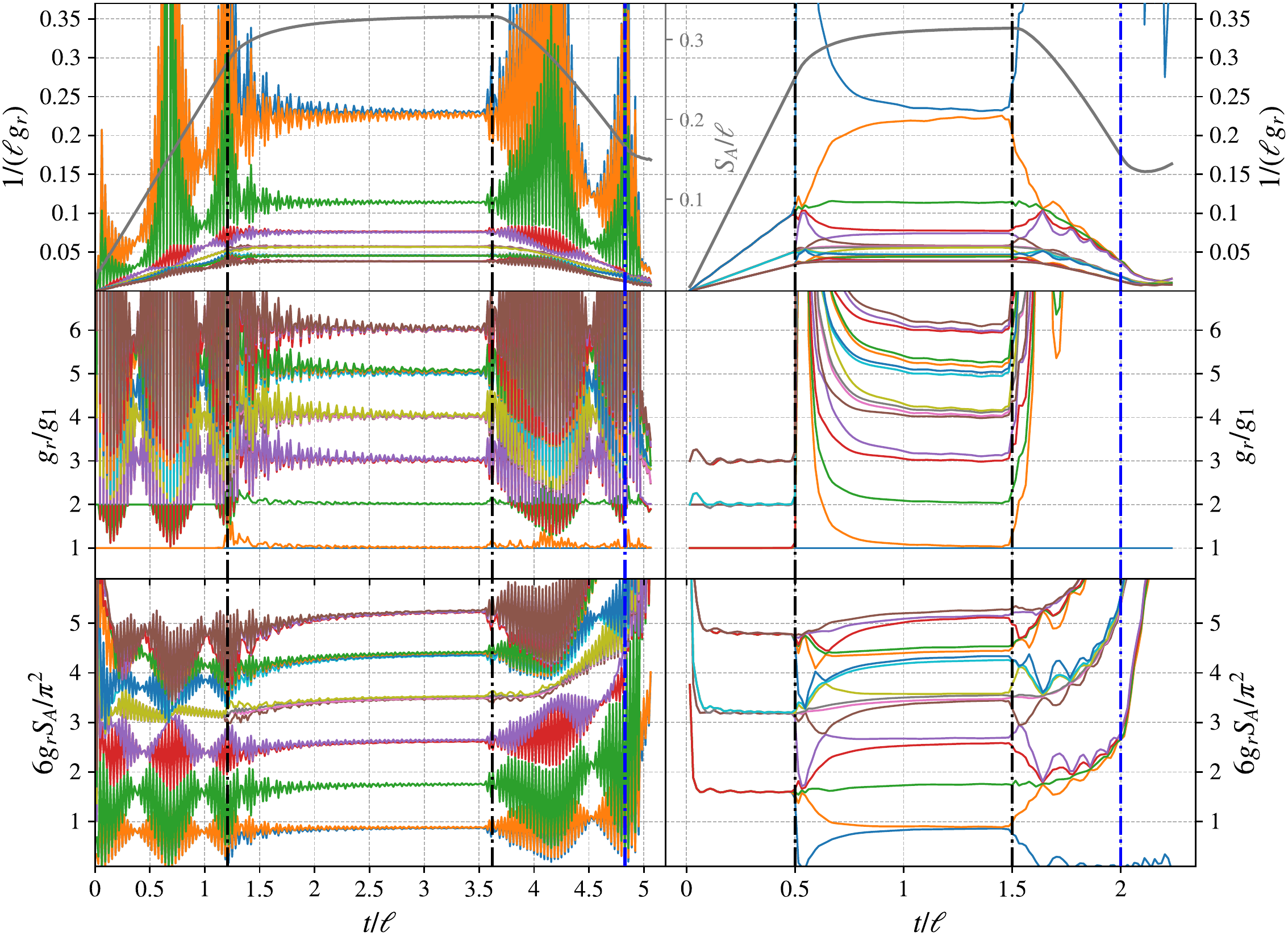}
\end{center}
\vspace{-.6cm}
\caption{(color online).
Time evolution of $(\ell g_r)^{-1}$ (top), $g_r/g_1$ (middle) and $g_r S_A$ (bottom)
after quenches such that $H$ and $H_0$ belong to different phases given by
$\theta_0 =  \pi/8 \to \theta = \pi/2 -\pi/8$ (left) 
and $\theta_0 = \pi/2 - \pi/8 \to \theta = \pi/8$ (right),
for an interval having $\ell=128$ sites in the chain with PBC having $L=512$ sites. 
}
\label{fig:AcrossPBC}
\end{figure}

\paragraph{Gapped evolution Hamiltonians.}
It is very instructive to perform the numerical analysis of the ES
discussed above for quenches where the post-quench Hamiltonian is gapped.
In Fig.\,\ref{fig:AcrossPBC} we report the results 
obtained for a typical quench across the QCP,
in both the directions. 
A major difference between the time evolutions of the ES 
after  these two protocols is the occurrence of singularities 
in the regimes I and III for the quench
from the paramagnetic phase to the ferromagnetic phase
(left panels).
In \cite{torlai2014}
it has been shown that the times at which 
the ES develops singularities in the regime I coincide with the 
times when the Loschmidt echo is singular,
identifying the quantum dynamical phase transition
\cite{heyl2013} (see \cite{heyl2019} for a recent review).

In the quench from the ferromagnetic phase 
to the paramagnetic phase (right panels), 
the evolutions are smooth and, surprisingly, 
in the regimes I and II for the ratios $g_r/g_1$ 
we observe plateaux corresponding to integer values.
It is remarkable that $g_r/g_1$ display 
plateaux at integer values in the time regime II
also for the quench 
from the paramagnetic phase to the ferromagnetic phase.
This feature, which has been highlighted for the quench 
at the QCP (see regime I in Fig.\,\ref{fig:ToCriticalPBC}),
could be attributed to the crossing of the QCP.
Indeed, in Fig.\,\ref{fig:InsidePhasePBC}, 
where the QCP is not crossed, these plateaux at integer values
for $g_r/g_1$ in the regime II are not observed. 
The collapse of the curves on the integer values in the regime II
of the quench 
from the paramagnetic phase to the ferromagnetic phase
improves as $|\theta - \theta_0|$ increases. 
According to the discussion reported above 
for the quenches at the QCP, these integer values 
could be related to two copies of 
the Ising BCFT on the annulus with CBC $a$ and $b_0$, 
like the regime I of the quench at the QCP 
(see Fig.\,\ref{fig:ToCriticalPBC}).
Analytic results supporting this observation are missing.
A possible interpretation could rely on the fact that,
being the evolution Hamiltonian gapped,
the correlation length is finite and therefore 
the entangling points 
can be considered independent from each other,
as in the regime I of the quench at the QCP.

 \begin{figure}[t!]
\vspace{-.0cm}
\begin{center}
\includegraphics[width=1.04\columnwidth]{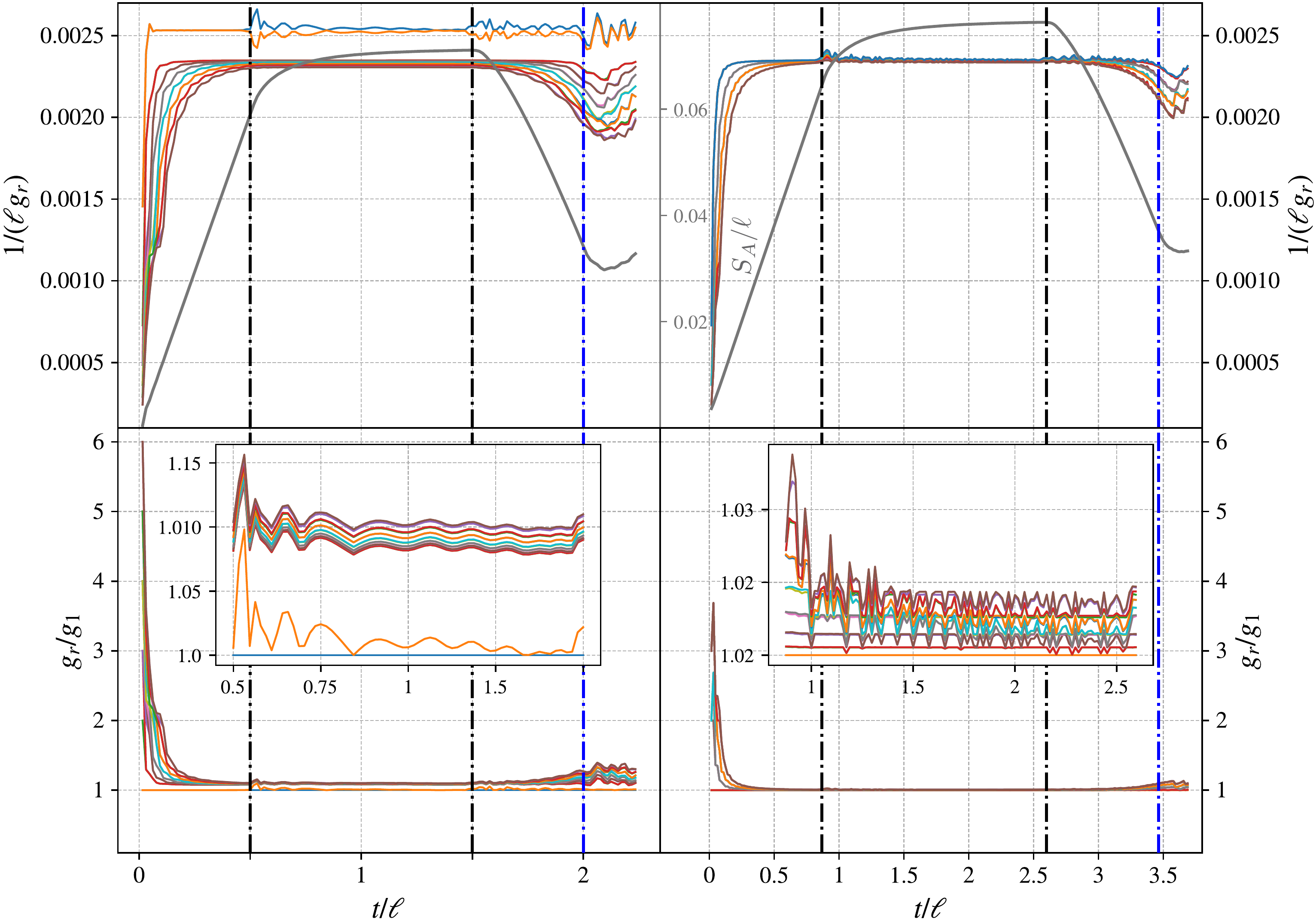}
\end{center}
\vspace{-.6cm}
\caption{(color online).
Time evolution of $(\ell g_r)^{-1}$ (top) and $g_r/g_1$ (bottom)
after the quench $\theta_0 =  \pi/12 \to \theta = \pi/4 -\pi/12$
within the paramagnetic phase (left)
and $\theta_0 = \pi/2-  \pi/12 \to \theta = \pi/4 + \pi/12$ 
within the ferromagnetic phase (right),
for an interval with $\ell=128$ sites in chain with PBC having $L=512$ sites. 
}
\label{fig:InsidePhasePBC}
\end{figure}

\paragraph{Summary and discussion.}
In finite TFIC with either PBC or OBC (reflecting boundaries), 
we have studied numerically the time evolution
of the gaps $g_r$ of the ES after a global quantum quench of the magnetic field,
when the subsystem is an interval.
We found that interesting information is provided by the ratios $g_r/g_1$,
which take constant values in various time regimes,
for the lower part of the ES.
In particular, 
when the post-quench Hamiltonian is critical,
the first thermalisation regime (regime II) 
is determined by the conformal spectrum
of the Ising BCFT on the annulus with the proper CBC.
Surprisingly, this feature is observed also 
for quenches across the QCP. 
Furthermore, these results are robust under reasonable
changes of the initial state. 
Our analysis leads to identify the proper CBC to adopt
in the BCFT approach to global quenches of \cite{calabrese2005c,calabrese2006a,calabrese2007,cardy2014,cardy2016g,cardy2016a}
that allow to reproduce the numerical results
(as done in \cite{lauchli2013a} at equilibrium in the ground state).

The numerical results reported here naturally require
further investigations in many directions. 
Repeating our analysis for finite 
open TFIC with different boundary conditions would improve  
the current understanding of the role of the boundaries
in the imaginary time path integral approach. 
It is worth extending 
our analysis to other spin chains
(e.g. XXZ and the Hubbard model),
both in the universality class of the TFIC and in different ones.
Reproducing our results through GGE methods would be very insightful.
The relation between the ES and the features of 
the dynamical quantum phase transitions 
deserves further, more quantitative, analysis.
Interesting results about the ES could be obtained also 
by considering directly the entanglement Hamiltonian
\cite{bisognano1975, peschel1999, peschel2003a, peschel2004a, peschel2009a, casini2011a, eisler2017a, eisler2019a, digiulio2019,fagotti2013b}
or bipartitions where $A$ is made by disjoint intervals
or spatially inhomogeneous lattice systems \cite{tonni2018}.
It is instuctive to study the time evolution of the ES after local quantum quenches \cite{j.suracea}.
Extending all these analysis to higher dimensions is also very important. 
Finally, we remark that our results naturally suggest that
experiments on entanglement spectroscopy 
of out-of-equilibrium correlated many-body quantum systems
could provide efficient methods to measure the critical exponents.

We are grateful to
Giuseppe Di Giulio, Glen Evenbly, Andreas Ludwig, Giuseppe Mussardo, 
Anatoli Polkovnikov and Marcos Rigol for useful discussions
and in particular to John Cardy for his comments on the draft.
JS is supported by the doctoral training partnership (DTP 2016-2017) of the University of Strathclyde.
LT is supported by the MINECO  RYC-2016-20594 fellowship and the MINECO PGC2018-095862-B-C22 grant.
We thank the organisers of the workshop 
{\it Entangle This IV: Chaos, Order and Qubits} (ICMAT, Madrid), 
where these results have been presented by LT. 
ET acknowledges Yukawa Institute for Theoretical Physics at Kyoto University, 
where part of this work has been done, during the workshop YITP-T-19-03 
{\it Quantum Information and String Theory 2019},
and Trieste Institute for Theoretical Quantum Technologies. 

\bibliographystyle{apsrev4-1}

\begin{thebibliography}{98}%
\makeatletter
\providecommand \@ifxundefined [1]{%
 \@ifx{#1\undefined}
}%
\providecommand \@ifnum [1]{%
 \ifnum #1\expandafter \@firstoftwo
 \else \expandafter \@secondoftwo
 \fi
}%
\providecommand \@ifx [1]{%
 \ifx #1\expandafter \@firstoftwo
 \else \expandafter \@secondoftwo
 \fi
}%
\providecommand \natexlab [1]{#1}%
\providecommand \enquote  [1]{``#1''}%
\providecommand \bibnamefont  [1]{#1}%
\providecommand \bibfnamefont [1]{#1}%
\providecommand \citenamefont [1]{#1}%
\providecommand \href@noop [0]{\@secondoftwo}%
\providecommand \href [0]{\begingroup \@sanitize@url \@href}%
\providecommand \@href[1]{\@@startlink{#1}\@@href}%
\providecommand \@@href[1]{\endgroup#1\@@endlink}%
\providecommand \@sanitize@url [0]{\catcode `\\12\catcode `\$12\catcode
  `\&12\catcode `\#12\catcode `\^12\catcode `\_12\catcode `\%12\relax}%
\providecommand \@@startlink[1]{}%
\providecommand \@@endlink[0]{}%
\providecommand \url  [0]{\begingroup\@sanitize@url \@url }%
\providecommand \@url [1]{\endgroup\@href {#1}{\urlprefix }}%
\providecommand \urlprefix  [0]{URL }%
\providecommand \Eprint [0]{\href }%
\providecommand \doibase [0]{http://dx.doi.org/}%
\providecommand \selectlanguage [0]{\@gobble}%
\providecommand \bibinfo  [0]{\@secondoftwo}%
\providecommand \bibfield  [0]{\@secondoftwo}%
\providecommand \translation [1]{[#1]}%
\providecommand \BibitemOpen [0]{}%
\providecommand \bibitemStop [0]{}%
\providecommand \bibitemNoStop [0]{.\EOS\space}%
\providecommand \EOS [0]{\spacefactor3000\relax}%
\providecommand \BibitemShut  [1]{\csname bibitem#1\endcsname}%
\let\auto@bib@innerbib\@empty
\bibitem [{\citenamefont {Hastings}(2007)}]{hastings2007}%
  \BibitemOpen
  \bibfield  {author} {\bibinfo {author} {\bibfnamefont {M.~B.}\ \bibnamefont
  {Hastings}},\ }\href {\doibase 10.1088/1742-5468/2007/08/P08024} {\bibfield
  {journal} {\bibinfo  {journal} {J. Stat. Mech.: Theory Exp.}\ }\textbf
  {\bibinfo {volume} {2007}},\ \bibinfo {pages} {P08024} (\bibinfo {year}
  {2007})}\BibitemShut {NoStop}%
\bibitem [{\citenamefont {Masanes}(2009)}]{masanes2009}%
  \BibitemOpen
  \bibfield  {author} {\bibinfo {author} {\bibfnamefont {L.}~\bibnamefont
  {Masanes}},\ }\href {\doibase 10.1103/PhysRevA.80.052104} {\bibfield
  {journal} {\bibinfo  {journal} {Phys. Rev. A}\ }\textbf {\bibinfo {volume}
  {80}},\ \bibinfo {pages} {052104} (\bibinfo {year} {2009})}\BibitemShut
  {NoStop}%
\bibitem [{\citenamefont {Amico}\ \emph {et~al.}(2008)\citenamefont {Amico},
  \citenamefont {Fazio}, \citenamefont {Osterloh},\ and\ \citenamefont
  {Vedral}}]{amico2008}%
  \BibitemOpen
  \bibfield  {author} {\bibinfo {author} {\bibfnamefont {L.}~\bibnamefont
  {Amico}}, \bibinfo {author} {\bibfnamefont {R.}~\bibnamefont {Fazio}},
  \bibinfo {author} {\bibfnamefont {A.}~\bibnamefont {Osterloh}}, \ and\
  \bibinfo {author} {\bibfnamefont {V.}~\bibnamefont {Vedral}},\ }\href
  {\doibase 10.1103/RevModPhys.80.517} {\bibfield  {journal} {\bibinfo
  {journal} {Rev. Mod. Phys.}\ }\textbf {\bibinfo {volume} {80}},\ \bibinfo
  {pages} {517} (\bibinfo {year} {2008})}\BibitemShut {NoStop}%
\bibitem [{\citenamefont {Bridgeman}\ and\ \citenamefont
  {Chubb}(2017)}]{bridgeman2017}%
  \BibitemOpen
  \bibfield  {author} {\bibinfo {author} {\bibfnamefont {J.~C.}\ \bibnamefont
  {Bridgeman}}\ and\ \bibinfo {author} {\bibfnamefont {C.~T.}\ \bibnamefont
  {Chubb}},\ }\href {\doibase 10.1088/1751-8121/aa6dc3} {\bibfield  {journal}
  {\bibinfo  {journal} {J. Phys. A: Math. Theor.}\ }\textbf {\bibinfo {volume}
  {50}},\ \bibinfo {pages} {223001} (\bibinfo {year} {2017})}\BibitemShut
  {NoStop}%
\bibitem [{\citenamefont {Or{\'u}s}(2014)}]{orus2014b}%
  \BibitemOpen
  \bibfield  {author} {\bibinfo {author} {\bibfnamefont {R.}~\bibnamefont
  {Or{\'u}s}},\ }\href {\doibase 10.1016/j.aop.2014.06.013} {\bibfield
  {journal} {\bibinfo  {journal} {Ann. Phys. (N. Y.)}\ }\textbf {\bibinfo
  {volume} {349}},\ \bibinfo {pages} {117} (\bibinfo {year}
  {2014})}\BibitemShut {NoStop}%
\bibitem [{\citenamefont {Ran}\ \emph {et~al.}(2017)\citenamefont {Ran},
  \citenamefont {Tirrito}, \citenamefont {Peng}, \citenamefont {Chen},
  \citenamefont {Tagliacozzo}, \citenamefont {Su},\ and\ \citenamefont
  {Lewenstein}}]{ran2017}%
  \BibitemOpen
  \bibfield  {author} {\bibinfo {author} {\bibfnamefont {S.-J.}\ \bibnamefont
  {Ran}}, \bibinfo {author} {\bibfnamefont {E.}~\bibnamefont {Tirrito}},
  \bibinfo {author} {\bibfnamefont {C.}~\bibnamefont {Peng}}, \bibinfo {author}
  {\bibfnamefont {X.}~\bibnamefont {Chen}}, \bibinfo {author} {\bibfnamefont
  {L.}~\bibnamefont {Tagliacozzo}}, \bibinfo {author} {\bibfnamefont
  {G.}~\bibnamefont {Su}}, \ and\ \bibinfo {author} {\bibfnamefont
  {M.}~\bibnamefont {Lewenstein}},\ }\href@noop {} {\  (\bibinfo {year}
  {2017})},\ \Eprint {http://arxiv.org/abs/1708.09213} {arXiv:1708.09213}
  \BibitemShut {NoStop}%
\bibitem [{\citenamefont {Pelissetto}\ and\ \citenamefont
  {Vicari}(2002)}]{pelissetto2002}%
  \BibitemOpen
  \bibfield  {author} {\bibinfo {author} {\bibfnamefont {A.}~\bibnamefont
  {Pelissetto}}\ and\ \bibinfo {author} {\bibfnamefont {E.}~\bibnamefont
  {Vicari}},\ }\href {\doibase 10.1016/S0370-1573(02)00219-3} {\bibfield
  {journal} {\bibinfo  {journal} {Phys. Rep.}\ }\textbf {\bibinfo {volume}
  {368}},\ \bibinfo {pages} {549} (\bibinfo {year} {2002})}\BibitemShut
  {NoStop}%
\bibitem [{\citenamefont {Belavin}\ \emph {et~al.}(1984)\citenamefont
  {Belavin}, \citenamefont {Polyakov},\ and\ \citenamefont
  {Zamolodchikov}}]{belavin1984}%
  \BibitemOpen
  \bibfield  {author} {\bibinfo {author} {\bibfnamefont {A.~A.}\ \bibnamefont
  {Belavin}}, \bibinfo {author} {\bibfnamefont {A.~M.}\ \bibnamefont
  {Polyakov}}, \ and\ \bibinfo {author} {\bibfnamefont {A.~B.}\ \bibnamefont
  {Zamolodchikov}},\ }\href {\doibase 10.1016/0550-3213(84)90052-X} {\bibfield
  {journal} {\bibinfo  {journal} {Nucl. Phys. B}\ }\textbf {\bibinfo {volume}
  {B241}},\ \bibinfo {pages} {333} (\bibinfo {year} {1984})}\BibitemShut
  {NoStop}%
\bibitem [{\citenamefont {Henkel}(1999)}]{henkel1999}%
  \BibitemOpen
  \bibfield  {author} {\bibinfo {author} {\bibfnamefont {M.}~\bibnamefont
  {Henkel}},\ }\href@noop {} {{\selectlanguage {en}\emph {\bibinfo {title}
  {Conformal {{Invariance}} and {{Critical Phenomena}}}}}},\ Theoretical and
  {{Mathematical Physics}}\ (\bibinfo  {publisher} {{Springer-Verlag}},\
  \bibinfo {address} {{Berlin Heidelberg}},\ \bibinfo {year}
  {1999})\BibitemShut {NoStop}%
\bibitem [{\citenamefont {Holzhey}\ \emph {et~al.}(1994)\citenamefont
  {Holzhey}, \citenamefont {Larsen},\ and\ \citenamefont
  {Wilczek}}]{holzhey1994}%
  \BibitemOpen
  \bibfield  {author} {\bibinfo {author} {\bibfnamefont {C.}~\bibnamefont
  {Holzhey}}, \bibinfo {author} {\bibfnamefont {F.}~\bibnamefont {Larsen}}, \
  and\ \bibinfo {author} {\bibfnamefont {F.}~\bibnamefont {Wilczek}},\ }\href
  {\doibase 10.1016/0550-3213(94)90402-2} {\bibfield  {journal} {\bibinfo
  {journal} {Nucl. Phys. B}\ }\textbf {\bibinfo {volume} {424}},\ \bibinfo
  {pages} {443} (\bibinfo {year} {1994})}\BibitemShut {NoStop}%
\bibitem [{\citenamefont {Calabrese}\ and\ \citenamefont
  {Cardy}(2004)}]{calabrese2004}%
  \BibitemOpen
  \bibfield  {author} {\bibinfo {author} {\bibfnamefont {P.}~\bibnamefont
  {Calabrese}}\ and\ \bibinfo {author} {\bibfnamefont {J.}~\bibnamefont
  {Cardy}},\ }\href {\doibase 10.1088/1742-5468/2004/06/P06002} {\bibfield
  {journal} {\bibinfo  {journal} {J. Stat. Mech.: Theory Exp.}\ }\textbf
  {\bibinfo {volume} {2004}},\ \bibinfo {pages} {P06002} (\bibinfo {year}
  {2004})}\BibitemShut {NoStop}%
\bibitem [{\citenamefont {Vidal}\ \emph {et~al.}(2003)\citenamefont {Vidal},
  \citenamefont {Latorre}, \citenamefont {Rico},\ and\ \citenamefont
  {Kitaev}}]{vidal2003a}%
  \BibitemOpen
  \bibfield  {author} {\bibinfo {author} {\bibfnamefont {G.}~\bibnamefont
  {Vidal}}, \bibinfo {author} {\bibfnamefont {J.~I.}\ \bibnamefont {Latorre}},
  \bibinfo {author} {\bibfnamefont {E.}~\bibnamefont {Rico}}, \ and\ \bibinfo
  {author} {\bibfnamefont {A.}~\bibnamefont {Kitaev}},\ }\href {\doibase
  10.1103/PhysRevLett.90.227902} {\bibfield  {journal} {\bibinfo  {journal}
  {Phys. Rev. Lett.}\ }\textbf {\bibinfo {volume} {90}},\ \bibinfo {pages}
  {227902} (\bibinfo {year} {2003})}\BibitemShut {NoStop}%
\bibitem [{\citenamefont {Callan}\ and\ \citenamefont
  {Wilczek}(1994)}]{callan1994b}%
  \BibitemOpen
  \bibfield  {author} {\bibinfo {author} {\bibfnamefont {C.}~\bibnamefont
  {Callan}}\ and\ \bibinfo {author} {\bibfnamefont {F.}~\bibnamefont
  {Wilczek}},\ }\href {\doibase 10.1016/0370-2693(94)91007-3} {\bibfield
  {journal} {\bibinfo  {journal} {Phys. Lett. B}\ }\textbf {\bibinfo {volume}
  {333}},\ \bibinfo {pages} {55} (\bibinfo {year} {1994})}\BibitemShut
  {NoStop}%
\bibitem [{\citenamefont {Calabrese}\ \emph {et~al.}(2009)\citenamefont
  {Calabrese}, \citenamefont {Cardy},\ and\ \citenamefont
  {Tonni}}]{calabrese2009b}%
  \BibitemOpen
  \bibfield  {author} {\bibinfo {author} {\bibfnamefont {P.}~\bibnamefont
  {Calabrese}}, \bibinfo {author} {\bibfnamefont {J.}~\bibnamefont {Cardy}}, \
  and\ \bibinfo {author} {\bibfnamefont {E.}~\bibnamefont {Tonni}},\ }\href
  {\doibase 10.1088/1742-5468/2009/11/P11001} {\bibfield  {journal} {\bibinfo
  {journal} {J. Stat. Mech.: Theory Exp.}\ }\textbf {\bibinfo {volume}
  {2009}},\ \bibinfo {pages} {P11001} (\bibinfo {year} {2009})}\BibitemShut
  {NoStop}%
\bibitem [{\citenamefont {Calabrese}\ \emph {et~al.}(2011)\citenamefont
  {Calabrese}, \citenamefont {Cardy},\ and\ \citenamefont
  {Tonni}}]{calabrese2011b}%
  \BibitemOpen
  \bibfield  {author} {\bibinfo {author} {\bibfnamefont {P.}~\bibnamefont
  {Calabrese}}, \bibinfo {author} {\bibfnamefont {J.}~\bibnamefont {Cardy}}, \
  and\ \bibinfo {author} {\bibfnamefont {E.}~\bibnamefont {Tonni}},\ }\href
  {\doibase 10.1088/1742-5468/2011/01/P01021} {\bibfield  {journal} {\bibinfo
  {journal} {J. Stat. Mech.: Theory Exp.}\ }\textbf {\bibinfo {volume}
  {2011}},\ \bibinfo {pages} {P01021} (\bibinfo {year} {2011})}\BibitemShut
  {NoStop}%
\bibitem [{\citenamefont {Alba}\ \emph {et~al.}(2011)\citenamefont {Alba},
  \citenamefont {Tagliacozzo},\ and\ \citenamefont {Calabrese}}]{alba2011}%
  \BibitemOpen
  \bibfield  {author} {\bibinfo {author} {\bibfnamefont {V.}~\bibnamefont
  {Alba}}, \bibinfo {author} {\bibfnamefont {L.}~\bibnamefont {Tagliacozzo}}, \
  and\ \bibinfo {author} {\bibfnamefont {P.}~\bibnamefont {Calabrese}},\ }\href
  {\doibase 10.1088/1742-5468/2011/06/P06012} {\bibfield  {journal} {\bibinfo
  {journal} {J. Stat. Mech.: Theory Exp.}\ }\textbf {\bibinfo {volume}
  {2011}},\ \bibinfo {pages} {P06012} (\bibinfo {year} {2011})}\BibitemShut
  {NoStop}%
\bibitem [{\citenamefont {Coser}\ \emph
  {et~al.}(2014{\natexlab{a}})\citenamefont {Coser}, \citenamefont
  {Tagliacozzo},\ and\ \citenamefont {Tonni}}]{coser2014a}%
  \BibitemOpen
  \bibfield  {author} {\bibinfo {author} {\bibfnamefont {A.}~\bibnamefont
  {Coser}}, \bibinfo {author} {\bibfnamefont {L.}~\bibnamefont {Tagliacozzo}},
  \ and\ \bibinfo {author} {\bibfnamefont {E.}~\bibnamefont {Tonni}},\ }\href
  {\doibase 10.1088/1742-5468/2014/01/P01008} {\bibfield  {journal} {\bibinfo
  {journal} {J. Stat. Mech.: Theory Exp.}\ }\textbf {\bibinfo {volume}
  {2014}},\ \bibinfo {pages} {P01008} (\bibinfo {year}
  {2014}{\natexlab{a}})}\BibitemShut {NoStop}%
\bibitem [{\citenamefont {De~Nobili}\ \emph {et~al.}(2015)\citenamefont
  {De~Nobili}, \citenamefont {Coser},\ and\ \citenamefont
  {Tonni}}]{denobili2015}%
  \BibitemOpen
  \bibfield  {author} {\bibinfo {author} {\bibfnamefont {C.}~\bibnamefont
  {De~Nobili}}, \bibinfo {author} {\bibfnamefont {A.}~\bibnamefont {Coser}}, \
  and\ \bibinfo {author} {\bibfnamefont {E.}~\bibnamefont {Tonni}},\ }\href
  {\doibase 10.1088/1742-5468/2015/06/P06021} {\bibfield  {journal} {\bibinfo
  {journal} {J. Stat. Mech.: Theory Exp.}\ }\textbf {\bibinfo {volume}
  {2015}},\ \bibinfo {pages} {P06021} (\bibinfo {year} {2015})}\BibitemShut
  {NoStop}%
\bibitem [{\citenamefont {Li}\ and\ \citenamefont {Haldane}(2008)}]{li2008}%
  \BibitemOpen
  \bibfield  {author} {\bibinfo {author} {\bibfnamefont {H.}~\bibnamefont
  {Li}}\ and\ \bibinfo {author} {\bibfnamefont {F.~D.~M.}\ \bibnamefont
  {Haldane}},\ }\href {\doibase 10.1103/PhysRevLett.101.010504} {\bibfield
  {journal} {\bibinfo  {journal} {Phys. Rev. Lett.}\ }\textbf {\bibinfo
  {volume} {101}},\ \bibinfo {pages} {010504} (\bibinfo {year}
  {2008})}\BibitemShut {NoStop}%
\bibitem [{\citenamefont {Peschel}\ and\ \citenamefont
  {Truong}(1987)}]{peschel1987}%
  \BibitemOpen
  \bibfield  {author} {\bibinfo {author} {\bibfnamefont {I.}~\bibnamefont
  {Peschel}}\ and\ \bibinfo {author} {\bibfnamefont {T.~T.}\ \bibnamefont
  {Truong}},\ }\href {\doibase 10.1007/BF01307296} {\bibfield  {journal}
  {\bibinfo  {journal} {Z. Physik B}\ }\textbf {\bibinfo {volume} {69}},\
  \bibinfo {pages} {385} (\bibinfo {year} {1987})}\BibitemShut {NoStop}%
\bibitem [{\citenamefont {Calabrese}\ and\ \citenamefont
  {Lefevre}(2008)}]{calabrese2008}%
  \BibitemOpen
  \bibfield  {author} {\bibinfo {author} {\bibfnamefont {P.}~\bibnamefont
  {Calabrese}}\ and\ \bibinfo {author} {\bibfnamefont {A.}~\bibnamefont
  {Lefevre}},\ }\href {\doibase 10.1103/PhysRevA.78.032329} {\bibfield
  {journal} {\bibinfo  {journal} {Phys. Rev. A}\ }\textbf {\bibinfo {volume}
  {78}},\ \bibinfo {pages} {032329} (\bibinfo {year} {2008})}\BibitemShut
  {NoStop}%
\bibitem [{\citenamefont {Poilblanc}(2010)}]{poilblanc2010}%
  \BibitemOpen
  \bibfield  {author} {\bibinfo {author} {\bibfnamefont {D.}~\bibnamefont
  {Poilblanc}},\ }\href {\doibase 10.1103/PhysRevLett.105.077202} {\bibfield
  {journal} {\bibinfo  {journal} {Phys. Rev. Lett.}\ }\textbf {\bibinfo
  {volume} {105}},\ \bibinfo {pages} {077202} (\bibinfo {year}
  {2010})}\BibitemShut {NoStop}%
\bibitem [{\citenamefont {Cirac}\ \emph {et~al.}(2011)\citenamefont {Cirac},
  \citenamefont {Poilblanc}, \citenamefont {Schuch},\ and\ \citenamefont
  {Verstraete}}]{cirac2011}%
  \BibitemOpen
  \bibfield  {author} {\bibinfo {author} {\bibfnamefont {J.~I.}\ \bibnamefont
  {Cirac}}, \bibinfo {author} {\bibfnamefont {D.}~\bibnamefont {Poilblanc}},
  \bibinfo {author} {\bibfnamefont {N.}~\bibnamefont {Schuch}}, \ and\ \bibinfo
  {author} {\bibfnamefont {F.}~\bibnamefont {Verstraete}},\ }\href {\doibase
  10.1103/PhysRevB.83.245134} {\bibfield  {journal} {\bibinfo  {journal} {Phys.
  Rev. B}\ }\textbf {\bibinfo {volume} {83}},\ \bibinfo {pages} {245134}
  (\bibinfo {year} {2011})}\BibitemShut {NoStop}%
\bibitem [{\citenamefont {Dubail}\ \emph {et~al.}(2012)\citenamefont {Dubail},
  \citenamefont {Read},\ and\ \citenamefont {Rezayi}}]{dubail2012}%
  \BibitemOpen
  \bibfield  {author} {\bibinfo {author} {\bibfnamefont {J.}~\bibnamefont
  {Dubail}}, \bibinfo {author} {\bibfnamefont {N.}~\bibnamefont {Read}}, \ and\
  \bibinfo {author} {\bibfnamefont {E.~H.}\ \bibnamefont {Rezayi}},\ }\href
  {\doibase 10.1103/PhysRevB.85.115321} {\bibfield  {journal} {\bibinfo
  {journal} {Phys. Rev. B}\ }\textbf {\bibinfo {volume} {85}},\ \bibinfo
  {pages} {115321} (\bibinfo {year} {2012})}\BibitemShut {NoStop}%
\bibitem [{\citenamefont {L{\"a}uchli}(2013)}]{lauchli2013a}%
  \BibitemOpen
  \bibfield  {author} {\bibinfo {author} {\bibfnamefont {A.~M.}\ \bibnamefont
  {L{\"a}uchli}},\ }\href@noop {} {\  (\bibinfo {year} {2013})},\ \Eprint
  {http://arxiv.org/abs/1303.0741} {arXiv:1303.0741} \BibitemShut {NoStop}%
\bibitem [{\citenamefont {Cardy}\ and\ \citenamefont
  {Tonni}(2016)}]{cardy2016a}%
  \BibitemOpen
  \bibfield  {author} {\bibinfo {author} {\bibfnamefont {J.}~\bibnamefont
  {Cardy}}\ and\ \bibinfo {author} {\bibfnamefont {E.}~\bibnamefont {Tonni}},\
  }\href {\doibase 10.1088/1742-5468/2016/12/123103} {\bibfield  {journal}
  {\bibinfo  {journal} {J. Stat. Mech.: Theory Exp.}\ }\textbf {\bibinfo
  {volume} {2016}},\ \bibinfo {pages} {123103} (\bibinfo {year}
  {2016})}\BibitemShut {NoStop}%
\bibitem [{\citenamefont {Cardy}(2011)}]{cardy2011}%
  \BibitemOpen
  \bibfield  {author} {\bibinfo {author} {\bibfnamefont {J.}~\bibnamefont
  {Cardy}},\ }\href {\doibase 10.1103/PhysRevLett.106.150404} {\bibfield
  {journal} {\bibinfo  {journal} {Phys. Rev. Lett.}\ }\textbf {\bibinfo
  {volume} {106}},\ \bibinfo {pages} {150404} (\bibinfo {year}
  {2011})}\BibitemShut {NoStop}%
\bibitem [{\citenamefont {Abanin}\ and\ \citenamefont
  {Demler}(2012)}]{abanin2012}%
  \BibitemOpen
  \bibfield  {author} {\bibinfo {author} {\bibfnamefont {D.~A.}\ \bibnamefont
  {Abanin}}\ and\ \bibinfo {author} {\bibfnamefont {E.}~\bibnamefont
  {Demler}},\ }\href {\doibase 10.1103/PhysRevLett.109.020504} {\bibfield
  {journal} {\bibinfo  {journal} {Phys. Rev. Lett.}\ }\textbf {\bibinfo
  {volume} {109}},\ \bibinfo {pages} {020504} (\bibinfo {year}
  {2012})}\BibitemShut {NoStop}%
\bibitem [{\citenamefont {Daley}\ \emph {et~al.}(2012)\citenamefont {Daley},
  \citenamefont {Pichler}, \citenamefont {Schachenmayer},\ and\ \citenamefont
  {Zoller}}]{daley2012}%
  \BibitemOpen
  \bibfield  {author} {\bibinfo {author} {\bibfnamefont {A.~J.}\ \bibnamefont
  {Daley}}, \bibinfo {author} {\bibfnamefont {H.}~\bibnamefont {Pichler}},
  \bibinfo {author} {\bibfnamefont {J.}~\bibnamefont {Schachenmayer}}, \ and\
  \bibinfo {author} {\bibfnamefont {P.}~\bibnamefont {Zoller}},\ }\href
  {\doibase 10.1103/PhysRevLett.109.020505} {\bibfield  {journal} {\bibinfo
  {journal} {Phys. Rev. Lett.}\ }\textbf {\bibinfo {volume} {109}},\ \bibinfo
  {pages} {020505} (\bibinfo {year} {2012})}\BibitemShut {NoStop}%
\bibitem [{\citenamefont {Islam}\ \emph {et~al.}(2015)\citenamefont {Islam},
  \citenamefont {Ma}, \citenamefont {Preiss}, \citenamefont {Eric~Tai},
  \citenamefont {Lukin}, \citenamefont {Rispoli},\ and\ \citenamefont
  {Greiner}}]{islam2015}%
  \BibitemOpen
  \bibfield  {author} {\bibinfo {author} {\bibfnamefont {R.}~\bibnamefont
  {Islam}}, \bibinfo {author} {\bibfnamefont {R.}~\bibnamefont {Ma}}, \bibinfo
  {author} {\bibfnamefont {P.~M.}\ \bibnamefont {Preiss}}, \bibinfo {author}
  {\bibfnamefont {M.}~\bibnamefont {Eric~Tai}}, \bibinfo {author}
  {\bibfnamefont {A.}~\bibnamefont {Lukin}}, \bibinfo {author} {\bibfnamefont
  {M.}~\bibnamefont {Rispoli}}, \ and\ \bibinfo {author} {\bibfnamefont
  {M.}~\bibnamefont {Greiner}},\ }\href {\doibase 10.1038/nature15750}
  {\bibfield  {journal} {\bibinfo  {journal} {Nature}\ }\textbf {\bibinfo
  {volume} {528}},\ \bibinfo {pages} {77} (\bibinfo {year} {2015})}\BibitemShut
  {NoStop}%
\bibitem [{\citenamefont {Hauke}\ \emph {et~al.}(2016)\citenamefont {Hauke},
  \citenamefont {Heyl}, \citenamefont {Tagliacozzo},\ and\ \citenamefont
  {Zoller}}]{hauke2016}%
  \BibitemOpen
  \bibfield  {author} {\bibinfo {author} {\bibfnamefont {P.}~\bibnamefont
  {Hauke}}, \bibinfo {author} {\bibfnamefont {M.}~\bibnamefont {Heyl}},
  \bibinfo {author} {\bibfnamefont {L.}~\bibnamefont {Tagliacozzo}}, \ and\
  \bibinfo {author} {\bibfnamefont {P.}~\bibnamefont {Zoller}},\ }\href
  {\doibase 10.1038/nphys3700} {\bibfield  {journal} {\bibinfo  {journal} {Nat.
  Phys.}\ }\textbf {\bibinfo {volume} {12}},\ \bibinfo {pages} {778} (\bibinfo
  {year} {2016})}\BibitemShut {NoStop}%
\bibitem [{\citenamefont {Pichler}\ \emph {et~al.}(2016)\citenamefont
  {Pichler}, \citenamefont {Zhu}, \citenamefont {Seif}, \citenamefont
  {Zoller},\ and\ \citenamefont {Hafezi}}]{pichler2016}%
  \BibitemOpen
  \bibfield  {author} {\bibinfo {author} {\bibfnamefont {H.}~\bibnamefont
  {Pichler}}, \bibinfo {author} {\bibfnamefont {G.}~\bibnamefont {Zhu}},
  \bibinfo {author} {\bibfnamefont {A.}~\bibnamefont {Seif}}, \bibinfo {author}
  {\bibfnamefont {P.}~\bibnamefont {Zoller}}, \ and\ \bibinfo {author}
  {\bibfnamefont {M.}~\bibnamefont {Hafezi}},\ }\href {\doibase
  10.1103/PhysRevX.6.041033} {\bibfield  {journal} {\bibinfo  {journal} {Phys.
  Rev. X}\ }\textbf {\bibinfo {volume} {6}},\ \bibinfo {pages} {041033}
  (\bibinfo {year} {2016})}\BibitemShut {NoStop}%
\bibitem [{\citenamefont {Kaufman}\ \emph {et~al.}(2016)\citenamefont
  {Kaufman}, \citenamefont {Tai}, \citenamefont {Lukin}, \citenamefont
  {Rispoli}, \citenamefont {Schittko}, \citenamefont {Preiss},\ and\
  \citenamefont {Greiner}}]{kaufman2016}%
  \BibitemOpen
  \bibfield  {author} {\bibinfo {author} {\bibfnamefont {A.~M.}\ \bibnamefont
  {Kaufman}}, \bibinfo {author} {\bibfnamefont {M.~E.}\ \bibnamefont {Tai}},
  \bibinfo {author} {\bibfnamefont {A.}~\bibnamefont {Lukin}}, \bibinfo
  {author} {\bibfnamefont {M.}~\bibnamefont {Rispoli}}, \bibinfo {author}
  {\bibfnamefont {R.}~\bibnamefont {Schittko}}, \bibinfo {author}
  {\bibfnamefont {P.~M.}\ \bibnamefont {Preiss}}, \ and\ \bibinfo {author}
  {\bibfnamefont {M.}~\bibnamefont {Greiner}},\ }\href {\doibase
  10.1126/science.aaf6725} {\bibfield  {journal} {\bibinfo  {journal}
  {Science}\ }\textbf {\bibinfo {volume} {353}},\ \bibinfo {pages} {794}
  (\bibinfo {year} {2016})}\BibitemShut {NoStop}%
\bibitem [{\citenamefont {Brydges}\ \emph {et~al.}(2019)\citenamefont
  {Brydges}, \citenamefont {Elben}, \citenamefont {Jurcevic}, \citenamefont
  {Vermersch}, \citenamefont {Maier}, \citenamefont {Lanyon}, \citenamefont
  {Zoller}, \citenamefont {Blatt},\ and\ \citenamefont {Roos}}]{brydges2019}%
  \BibitemOpen
  \bibfield  {author} {\bibinfo {author} {\bibfnamefont {T.}~\bibnamefont
  {Brydges}}, \bibinfo {author} {\bibfnamefont {A.}~\bibnamefont {Elben}},
  \bibinfo {author} {\bibfnamefont {P.}~\bibnamefont {Jurcevic}}, \bibinfo
  {author} {\bibfnamefont {B.}~\bibnamefont {Vermersch}}, \bibinfo {author}
  {\bibfnamefont {C.}~\bibnamefont {Maier}}, \bibinfo {author} {\bibfnamefont
  {B.~P.}\ \bibnamefont {Lanyon}}, \bibinfo {author} {\bibfnamefont
  {P.}~\bibnamefont {Zoller}}, \bibinfo {author} {\bibfnamefont
  {R.}~\bibnamefont {Blatt}}, \ and\ \bibinfo {author} {\bibfnamefont {C.~F.}\
  \bibnamefont {Roos}},\ }\href {\doibase 10.1126/science.aau4963} {\bibfield
  {journal} {\bibinfo  {journal} {Science}\ }\textbf {\bibinfo {volume}
  {364}},\ \bibinfo {pages} {260} (\bibinfo {year} {2019})}\BibitemShut
  {NoStop}%
\bibitem [{\citenamefont {Lukin}\ \emph {et~al.}(2019)\citenamefont {Lukin},
  \citenamefont {Rispoli}, \citenamefont {Schittko}, \citenamefont {Tai},
  \citenamefont {Kaufman}, \citenamefont {Choi}, \citenamefont {Khemani},
  \citenamefont {L{\'e}onard},\ and\ \citenamefont {Greiner}}]{lukin2019}%
  \BibitemOpen
  \bibfield  {author} {\bibinfo {author} {\bibfnamefont {A.}~\bibnamefont
  {Lukin}}, \bibinfo {author} {\bibfnamefont {M.}~\bibnamefont {Rispoli}},
  \bibinfo {author} {\bibfnamefont {R.}~\bibnamefont {Schittko}}, \bibinfo
  {author} {\bibfnamefont {M.~E.}\ \bibnamefont {Tai}}, \bibinfo {author}
  {\bibfnamefont {A.~M.}\ \bibnamefont {Kaufman}}, \bibinfo {author}
  {\bibfnamefont {S.}~\bibnamefont {Choi}}, \bibinfo {author} {\bibfnamefont
  {V.}~\bibnamefont {Khemani}}, \bibinfo {author} {\bibfnamefont
  {J.}~\bibnamefont {L{\'e}onard}}, \ and\ \bibinfo {author} {\bibfnamefont
  {M.}~\bibnamefont {Greiner}},\ }\href {\doibase 10.1126/science.aau0818}
  {\bibfield  {journal} {\bibinfo  {journal} {Science}\ }\textbf {\bibinfo
  {volume} {364}},\ \bibinfo {pages} {256} (\bibinfo {year}
  {2019})}\BibitemShut {NoStop}%
\bibitem [{\citenamefont {Calabrese}\ and\ \citenamefont
  {Cardy}(2005)}]{calabrese2005c}%
  \BibitemOpen
  \bibfield  {author} {\bibinfo {author} {\bibfnamefont {P.}~\bibnamefont
  {Calabrese}}\ and\ \bibinfo {author} {\bibfnamefont {J.}~\bibnamefont
  {Cardy}},\ }\href {\doibase 10.1088/1742-5468/2005/04/P04010} {\bibfield
  {journal} {\bibinfo  {journal} {J. Stat. Mech.: Theory Exp.}\ }\textbf
  {\bibinfo {volume} {2005}},\ \bibinfo {pages} {P04010} (\bibinfo {year}
  {2005})}\BibitemShut {NoStop}%
\bibitem [{\citenamefont {De~Chiara}\ \emph {et~al.}(2006)\citenamefont
  {De~Chiara}, \citenamefont {Montangero}, \citenamefont {Calabrese},\ and\
  \citenamefont {Fazio}}]{dechiara2006}%
  \BibitemOpen
  \bibfield  {author} {\bibinfo {author} {\bibfnamefont {G.}~\bibnamefont
  {De~Chiara}}, \bibinfo {author} {\bibfnamefont {S.}~\bibnamefont
  {Montangero}}, \bibinfo {author} {\bibfnamefont {P.}~\bibnamefont
  {Calabrese}}, \ and\ \bibinfo {author} {\bibfnamefont {R.}~\bibnamefont
  {Fazio}},\ }\href {\doibase 10.1088/1742-5468/2006/03/P03001} {\bibfield
  {journal} {\bibinfo  {journal} {J. Stat. Mech.: Theory Exp.}\ }\textbf
  {\bibinfo {volume} {2006}},\ \bibinfo {pages} {P03001} (\bibinfo {year}
  {2006})}\BibitemShut {NoStop}%
\bibitem [{\citenamefont {Coser}\ \emph
  {et~al.}(2014{\natexlab{b}})\citenamefont {Coser}, \citenamefont {Tonni},\
  and\ \citenamefont {Calabrese}}]{coser2014b}%
  \BibitemOpen
  \bibfield  {author} {\bibinfo {author} {\bibfnamefont {A.}~\bibnamefont
  {Coser}}, \bibinfo {author} {\bibfnamefont {E.}~\bibnamefont {Tonni}}, \ and\
  \bibinfo {author} {\bibfnamefont {P.}~\bibnamefont {Calabrese}},\ }\href
  {\doibase 10.1088/1742-5468/2014/12/P12017} {\bibfield  {journal} {\bibinfo
  {journal} {J. Stat. Mech.: Theory Exp.}\ }\textbf {\bibinfo {volume}
  {2014}},\ \bibinfo {pages} {P12017} (\bibinfo {year}
  {2014}{\natexlab{b}})}\BibitemShut {NoStop}%
\bibitem [{\citenamefont {Nahum}\ \emph {et~al.}(2017)\citenamefont {Nahum},
  \citenamefont {Ruhman}, \citenamefont {Vijay},\ and\ \citenamefont
  {Haah}}]{nahum2017}%
  \BibitemOpen
  \bibfield  {author} {\bibinfo {author} {\bibfnamefont {A.}~\bibnamefont
  {Nahum}}, \bibinfo {author} {\bibfnamefont {J.}~\bibnamefont {Ruhman}},
  \bibinfo {author} {\bibfnamefont {S.}~\bibnamefont {Vijay}}, \ and\ \bibinfo
  {author} {\bibfnamefont {J.}~\bibnamefont {Haah}},\ }\href {\doibase
  10.1103/PhysRevX.7.031016} {\bibfield  {journal} {\bibinfo  {journal} {Phys.
  Rev. X}\ }\textbf {\bibinfo {volume} {7}},\ \bibinfo {pages} {031016}
  (\bibinfo {year} {2017})}\BibitemShut {NoStop}%
\bibitem [{\citenamefont {Vidal}(2004)}]{vidal2004}%
  \BibitemOpen
  \bibfield  {author} {\bibinfo {author} {\bibfnamefont {G.}~\bibnamefont
  {Vidal}},\ }\href {\doibase 10.1103/PhysRevLett.93.040502} {\bibfield
  {journal} {\bibinfo  {journal} {Phys. Rev. Lett.}\ }\textbf {\bibinfo
  {volume} {93}},\ \bibinfo {pages} {040502} (\bibinfo {year}
  {2004})}\BibitemShut {NoStop}%
\bibitem [{\citenamefont {White}\ and\ \citenamefont
  {Feiguin}(2004)}]{white2004}%
  \BibitemOpen
  \bibfield  {author} {\bibinfo {author} {\bibfnamefont {S.~R.}\ \bibnamefont
  {White}}\ and\ \bibinfo {author} {\bibfnamefont {A.~E.}\ \bibnamefont
  {Feiguin}},\ }\href {\doibase 10.1103/PhysRevLett.93.076401} {\bibfield
  {journal} {\bibinfo  {journal} {Phys. Rev. Lett.}\ }\textbf {\bibinfo
  {volume} {93}},\ \bibinfo {pages} {076401} (\bibinfo {year}
  {2004})}\BibitemShut {NoStop}%
\bibitem [{\citenamefont {Daley}\ \emph {et~al.}(2004)\citenamefont {Daley},
  \citenamefont {Kollath}, \citenamefont {Schollw{\"o}ck},\ and\ \citenamefont
  {Vidal}}]{daley2004}%
  \BibitemOpen
  \bibfield  {author} {\bibinfo {author} {\bibfnamefont {A.~J.}\ \bibnamefont
  {Daley}}, \bibinfo {author} {\bibfnamefont {C.}~\bibnamefont {Kollath}},
  \bibinfo {author} {\bibfnamefont {U.}~\bibnamefont {Schollw{\"o}ck}}, \ and\
  \bibinfo {author} {\bibfnamefont {G.}~\bibnamefont {Vidal}},\ }\href
  {\doibase 10.1088/1742-5468/2004/04/P04005} {\bibfield  {journal} {\bibinfo
  {journal} {J. Stat. Mech.: Theory Exp.}\ }\textbf {\bibinfo {volume}
  {2004}},\ \bibinfo {pages} {P04005} (\bibinfo {year} {2004})}\BibitemShut
  {NoStop}%
\bibitem [{\citenamefont {Paeckel}\ \emph {et~al.}(2019)\citenamefont
  {Paeckel}, \citenamefont {K{\"o}hler}, \citenamefont {Swoboda}, \citenamefont
  {Manmana}, \citenamefont {Schollw{\"o}ck},\ and\ \citenamefont
  {Hubig}}]{paeckel2019a}%
  \BibitemOpen
  \bibfield  {author} {\bibinfo {author} {\bibfnamefont {S.}~\bibnamefont
  {Paeckel}}, \bibinfo {author} {\bibfnamefont {T.}~\bibnamefont {K{\"o}hler}},
  \bibinfo {author} {\bibfnamefont {A.}~\bibnamefont {Swoboda}}, \bibinfo
  {author} {\bibfnamefont {S.~R.}\ \bibnamefont {Manmana}}, \bibinfo {author}
  {\bibfnamefont {U.}~\bibnamefont {Schollw{\"o}ck}}, \ and\ \bibinfo {author}
  {\bibfnamefont {C.}~\bibnamefont {Hubig}},\ }\href@noop {} {\  (\bibinfo
  {year} {2019})},\ \Eprint {http://arxiv.org/abs/1901.05824}
  {arXiv:1901.05824} \BibitemShut {NoStop}%
\bibitem [{\citenamefont {Leviatan}\ \emph {et~al.}(2017)\citenamefont
  {Leviatan}, \citenamefont {Pollmann}, \citenamefont {Bardarson},
  \citenamefont {Huse},\ and\ \citenamefont {Altman}}]{leviatan2017a}%
  \BibitemOpen
  \bibfield  {author} {\bibinfo {author} {\bibfnamefont {E.}~\bibnamefont
  {Leviatan}}, \bibinfo {author} {\bibfnamefont {F.}~\bibnamefont {Pollmann}},
  \bibinfo {author} {\bibfnamefont {J.~H.}\ \bibnamefont {Bardarson}}, \bibinfo
  {author} {\bibfnamefont {D.~A.}\ \bibnamefont {Huse}}, \ and\ \bibinfo
  {author} {\bibfnamefont {E.}~\bibnamefont {Altman}},\ }\href@noop {} {\
  (\bibinfo {year} {2017})},\ \Eprint {http://arxiv.org/abs/1702.08894}
  {arXiv:1702.08894} \BibitemShut {NoStop}%
\bibitem [{\citenamefont {White}\ \emph {et~al.}(2017)\citenamefont {White},
  \citenamefont {Zaletel}, \citenamefont {Mong},\ and\ \citenamefont
  {Refael}}]{white2017}%
  \BibitemOpen
  \bibfield  {author} {\bibinfo {author} {\bibfnamefont {C.~D.}\ \bibnamefont
  {White}}, \bibinfo {author} {\bibfnamefont {M.}~\bibnamefont {Zaletel}},
  \bibinfo {author} {\bibfnamefont {R.~S.~K.}\ \bibnamefont {Mong}}, \ and\
  \bibinfo {author} {\bibfnamefont {G.}~\bibnamefont {Refael}},\ }\href@noop {}
  {\  (\bibinfo {year} {2017})},\ \Eprint {http://arxiv.org/abs/1707.01506}
  {arXiv:1707.01506} \BibitemShut {NoStop}%
\bibitem [{\citenamefont {Surace}\ \emph {et~al.}(2019)\citenamefont {Surace},
  \citenamefont {Piani},\ and\ \citenamefont {Tagliacozzo}}]{surace2019}%
  \BibitemOpen
  \bibfield  {author} {\bibinfo {author} {\bibfnamefont {J.}~\bibnamefont
  {Surace}}, \bibinfo {author} {\bibfnamefont {M.}~\bibnamefont {Piani}}, \
  and\ \bibinfo {author} {\bibfnamefont {L.}~\bibnamefont {Tagliacozzo}},\
  }\href {\doibase 10.1103/PhysRevB.99.235115} {\bibfield  {journal} {\bibinfo
  {journal} {Phys. Rev. B}\ }\textbf {\bibinfo {volume} {99}},\ \bibinfo
  {pages} {235115} (\bibinfo {year} {2019})}\BibitemShut {NoStop}%
\bibitem [{\citenamefont {Rams}\ and\ \citenamefont {Zwolak}(2019)}]{rams2019}%
  \BibitemOpen
  \bibfield  {author} {\bibinfo {author} {\bibfnamefont {M.~M.}\ \bibnamefont
  {Rams}}\ and\ \bibinfo {author} {\bibfnamefont {M.}~\bibnamefont {Zwolak}},\
  }\href@noop {} {\  (\bibinfo {year} {2019})},\ \Eprint
  {http://arxiv.org/abs/1904.12793} {arXiv:1904.12793} \BibitemShut {NoStop}%
\bibitem [{\citenamefont {Krumnow}\ \emph {et~al.}(2019)\citenamefont
  {Krumnow}, \citenamefont {Eisert},\ and\ \citenamefont
  {Legeza}}]{krumnow2019}%
  \BibitemOpen
  \bibfield  {author} {\bibinfo {author} {\bibfnamefont {C.}~\bibnamefont
  {Krumnow}}, \bibinfo {author} {\bibfnamefont {J.}~\bibnamefont {Eisert}}, \
  and\ \bibinfo {author} {\bibfnamefont {{\"O}.}~\bibnamefont {Legeza}},\
  }\href@noop {} {\  (\bibinfo {year} {2019})},\ \Eprint
  {http://arxiv.org/abs/1904.11999} {arXiv:1904.11999} \BibitemShut {NoStop}%
\bibitem [{\citenamefont {Fausti}\ \emph {et~al.}(2011)\citenamefont {Fausti},
  \citenamefont {Tobey}, \citenamefont {Dean}, \citenamefont {Kaiser},
  \citenamefont {Dienst}, \citenamefont {Hoffmann}, \citenamefont {Pyon},
  \citenamefont {Takayama}, \citenamefont {Takagi},\ and\ \citenamefont
  {Cavalleri}}]{fausti2011}%
  \BibitemOpen
  \bibfield  {author} {\bibinfo {author} {\bibfnamefont {D.}~\bibnamefont
  {Fausti}}, \bibinfo {author} {\bibfnamefont {R.~I.}\ \bibnamefont {Tobey}},
  \bibinfo {author} {\bibfnamefont {N.}~\bibnamefont {Dean}}, \bibinfo {author}
  {\bibfnamefont {S.}~\bibnamefont {Kaiser}}, \bibinfo {author} {\bibfnamefont
  {A.}~\bibnamefont {Dienst}}, \bibinfo {author} {\bibfnamefont {M.~C.}\
  \bibnamefont {Hoffmann}}, \bibinfo {author} {\bibfnamefont {S.}~\bibnamefont
  {Pyon}}, \bibinfo {author} {\bibfnamefont {T.}~\bibnamefont {Takayama}},
  \bibinfo {author} {\bibfnamefont {H.}~\bibnamefont {Takagi}}, \ and\ \bibinfo
  {author} {\bibfnamefont {A.}~\bibnamefont {Cavalleri}},\ }\href {\doibase
  10.1126/science.1197294} {\bibfield  {journal} {\bibinfo  {journal}
  {Science}\ }\textbf {\bibinfo {volume} {331}},\ \bibinfo {pages} {189}
  (\bibinfo {year} {2011})}\BibitemShut {NoStop}%
\bibitem [{\citenamefont {Abanin}\ and\ \citenamefont
  {Papi{\'c}}(2017)}]{abanin2017}%
  \BibitemOpen
  \bibfield  {author} {\bibinfo {author} {\bibfnamefont {D.~A.}\ \bibnamefont
  {Abanin}}\ and\ \bibinfo {author} {\bibfnamefont {Z.}~\bibnamefont
  {Papi{\'c}}},\ }\href {\doibase 10.1002/andp.201700169} {\bibfield  {journal}
  {\bibinfo  {journal} {Annalen der Physik}\ }\textbf {\bibinfo {volume}
  {529}},\ \bibinfo {pages} {1700169} (\bibinfo {year} {2017})}\BibitemShut
  {NoStop}%
\bibitem [{\citenamefont {Alet}\ and\ \citenamefont
  {Laflorencie}(2018)}]{alet2018a}%
  \BibitemOpen
  \bibfield  {author} {\bibinfo {author} {\bibfnamefont {F.}~\bibnamefont
  {Alet}}\ and\ \bibinfo {author} {\bibfnamefont {N.}~\bibnamefont
  {Laflorencie}},\ }\href {\doibase 10.1016/j.crhy.2018.03.003} {\bibfield
  {journal} {\bibinfo  {journal} {Comptes Rendus Physique}\ }\textbf {\bibinfo
  {volume} {19}},\ \bibinfo {pages} {498} (\bibinfo {year} {2018})}\BibitemShut
  {NoStop}%
\bibitem [{\citenamefont {Rigol}\ \emph {et~al.}(2008)\citenamefont {Rigol},
  \citenamefont {Dunjko},\ and\ \citenamefont {Olshanii}}]{rigol2008}%
  \BibitemOpen
  \bibfield  {author} {\bibinfo {author} {\bibfnamefont {M.}~\bibnamefont
  {Rigol}}, \bibinfo {author} {\bibfnamefont {V.}~\bibnamefont {Dunjko}}, \
  and\ \bibinfo {author} {\bibfnamefont {M.}~\bibnamefont {Olshanii}},\ }\href
  {\doibase 10.1038/nature06838} {\bibfield  {journal} {\bibinfo  {journal}
  {Nature}\ }\textbf {\bibinfo {volume} {452}},\ \bibinfo {pages} {854}
  (\bibinfo {year} {2008})}\BibitemShut {NoStop}%
\bibitem [{\citenamefont {Polkovnikov}\ \emph {et~al.}(2011)\citenamefont
  {Polkovnikov}, \citenamefont {Sengupta}, \citenamefont {Silva},\ and\
  \citenamefont {Vengalattore}}]{polkovnikov2011}%
  \BibitemOpen
  \bibfield  {author} {\bibinfo {author} {\bibfnamefont {A.}~\bibnamefont
  {Polkovnikov}}, \bibinfo {author} {\bibfnamefont {K.}~\bibnamefont
  {Sengupta}}, \bibinfo {author} {\bibfnamefont {A.}~\bibnamefont {Silva}}, \
  and\ \bibinfo {author} {\bibfnamefont {M.}~\bibnamefont {Vengalattore}},\
  }\href {\doibase 10.1103/RevModPhys.83.863} {\bibfield  {journal} {\bibinfo
  {journal} {Rev. Mod. Phys.}\ }\textbf {\bibinfo {volume} {83}},\ \bibinfo
  {pages} {863} (\bibinfo {year} {2011})}\BibitemShut {NoStop}%
\bibitem [{\citenamefont {Eisert}\ \emph {et~al.}(2015)\citenamefont {Eisert},
  \citenamefont {Friesdorf},\ and\ \citenamefont {Gogolin}}]{eisert2015}%
  \BibitemOpen
  \bibfield  {author} {\bibinfo {author} {\bibfnamefont {J.}~\bibnamefont
  {Eisert}}, \bibinfo {author} {\bibfnamefont {M.}~\bibnamefont {Friesdorf}}, \
  and\ \bibinfo {author} {\bibfnamefont {C.}~\bibnamefont {Gogolin}},\ }\href
  {\doibase 10.1038/nphys3215} {\bibfield  {journal} {\bibinfo  {journal} {Nat.
  Phys.}\ }\textbf {\bibinfo {volume} {11}},\ \bibinfo {pages} {124} (\bibinfo
  {year} {2015})}\BibitemShut {NoStop}%
\bibitem [{\citenamefont {D'Alessio}\ \emph {et~al.}(2016)\citenamefont
  {D'Alessio}, \citenamefont {Kafri}, \citenamefont {Polkovnikov},\ and\
  \citenamefont {Rigol}}]{dalessio2016}%
  \BibitemOpen
  \bibfield  {author} {\bibinfo {author} {\bibfnamefont {L.}~\bibnamefont
  {D'Alessio}}, \bibinfo {author} {\bibfnamefont {Y.}~\bibnamefont {Kafri}},
  \bibinfo {author} {\bibfnamefont {A.}~\bibnamefont {Polkovnikov}}, \ and\
  \bibinfo {author} {\bibfnamefont {M.}~\bibnamefont {Rigol}},\ }\href
  {\doibase 10.1080/00018732.2016.1198134} {\bibfield  {journal} {\bibinfo
  {journal} {Adv. Phys.}\ }\textbf {\bibinfo {volume} {65}},\ \bibinfo {pages}
  {239} (\bibinfo {year} {2016})}\BibitemShut {NoStop}%
\bibitem [{\citenamefont {Pr{\"u}fer}\ \emph {et~al.}(2018)\citenamefont
  {Pr{\"u}fer}, \citenamefont {Kunkel}, \citenamefont {Strobel}, \citenamefont
  {Lannig}, \citenamefont {Linnemann}, \citenamefont {Schmied}, \citenamefont
  {Berges}, \citenamefont {Gasenzer},\ and\ \citenamefont
  {Oberthaler}}]{prufer2018a}%
  \BibitemOpen
  \bibfield  {author} {\bibinfo {author} {\bibfnamefont {M.}~\bibnamefont
  {Pr{\"u}fer}}, \bibinfo {author} {\bibfnamefont {P.}~\bibnamefont {Kunkel}},
  \bibinfo {author} {\bibfnamefont {H.}~\bibnamefont {Strobel}}, \bibinfo
  {author} {\bibfnamefont {S.}~\bibnamefont {Lannig}}, \bibinfo {author}
  {\bibfnamefont {D.}~\bibnamefont {Linnemann}}, \bibinfo {author}
  {\bibfnamefont {C.-M.}\ \bibnamefont {Schmied}}, \bibinfo {author}
  {\bibfnamefont {J.}~\bibnamefont {Berges}}, \bibinfo {author} {\bibfnamefont
  {T.}~\bibnamefont {Gasenzer}}, \ and\ \bibinfo {author} {\bibfnamefont
  {M.~K.}\ \bibnamefont {Oberthaler}},\ }\href {\doibase
  10.1038/s41586-018-0659-0} {\bibfield  {journal} {\bibinfo  {journal}
  {Nature}\ }\textbf {\bibinfo {volume} {563}},\ \bibinfo {pages} {217}
  (\bibinfo {year} {2018})}\BibitemShut {NoStop}%
\bibitem [{\citenamefont {Heyl}\ \emph {et~al.}(2013)\citenamefont {Heyl},
  \citenamefont {Polkovnikov},\ and\ \citenamefont {Kehrein}}]{heyl2013}%
  \BibitemOpen
  \bibfield  {author} {\bibinfo {author} {\bibfnamefont {M.}~\bibnamefont
  {Heyl}}, \bibinfo {author} {\bibfnamefont {A.}~\bibnamefont {Polkovnikov}}, \
  and\ \bibinfo {author} {\bibfnamefont {S.}~\bibnamefont {Kehrein}},\ }\href
  {\doibase 10.1103/PhysRevLett.110.135704} {\bibfield  {journal} {\bibinfo
  {journal} {Phys. Rev. Lett.}\ }\textbf {\bibinfo {volume} {110}},\ \bibinfo
  {pages} {135704} (\bibinfo {year} {2013})}\BibitemShut {NoStop}%
\bibitem [{\citenamefont {{\v Z}unkovi{\v c}}\ \emph
  {et~al.}(2018)\citenamefont {{\v Z}unkovi{\v c}}, \citenamefont {Heyl},
  \citenamefont {Knap},\ and\ \citenamefont {Silva}}]{zunkovic2018}%
  \BibitemOpen
  \bibfield  {author} {\bibinfo {author} {\bibfnamefont {B.}~\bibnamefont {{\v
  Z}unkovi{\v c}}}, \bibinfo {author} {\bibfnamefont {M.}~\bibnamefont {Heyl}},
  \bibinfo {author} {\bibfnamefont {M.}~\bibnamefont {Knap}}, \ and\ \bibinfo
  {author} {\bibfnamefont {A.}~\bibnamefont {Silva}},\ }\href {\doibase
  10.1103/PhysRevLett.120.130601} {\bibfield  {journal} {\bibinfo  {journal}
  {Phys. Rev. Lett.}\ }\textbf {\bibinfo {volume} {120}},\ \bibinfo {pages}
  {130601} (\bibinfo {year} {2018})}\BibitemShut {NoStop}%
\bibitem [{\citenamefont {Heyl}(2019)}]{heyl2019}%
  \BibitemOpen
  \bibfield  {author} {\bibinfo {author} {\bibfnamefont {M.}~\bibnamefont
  {Heyl}},\ }\href {\doibase 10.1209/0295-5075/125/26001} {\bibfield  {journal}
  {\bibinfo  {journal} {EPL}\ }\textbf {\bibinfo {volume} {125}},\ \bibinfo
  {pages} {26001} (\bibinfo {year} {2019})}\BibitemShut {NoStop}%
\bibitem [{\citenamefont {Jordan}\ and\ \citenamefont
  {Wigner}(1928)}]{jordan1928}%
  \BibitemOpen
  \bibfield  {author} {\bibinfo {author} {\bibfnamefont {P.}~\bibnamefont
  {Jordan}}\ and\ \bibinfo {author} {\bibfnamefont {E.}~\bibnamefont
  {Wigner}},\ }\href {\doibase 10.1007/BF01331938} {\bibfield  {journal}
  {\bibinfo  {journal} {Z. Phys.}\ }\textbf {\bibinfo {volume} {47}},\ \bibinfo
  {pages} {631} (\bibinfo {year} {1928})}\BibitemShut {NoStop}%
\bibitem [{\citenamefont {Lieb}\ \emph {et~al.}(1961)\citenamefont {Lieb},
  \citenamefont {Schultz},\ and\ \citenamefont {Mattis}}]{lieb1961}%
  \BibitemOpen
  \bibfield  {author} {\bibinfo {author} {\bibfnamefont {E.}~\bibnamefont
  {Lieb}}, \bibinfo {author} {\bibfnamefont {T.}~\bibnamefont {Schultz}}, \
  and\ \bibinfo {author} {\bibfnamefont {D.}~\bibnamefont {Mattis}},\ }\href
  {\doibase 10.1016/0003-4916(61)90115-4} {\bibfield  {journal} {\bibinfo
  {journal} {Ann. Phys. (N. Y.)}\ }\textbf {\bibinfo {volume} {16}},\ \bibinfo
  {pages} {407} (\bibinfo {year} {1961})}\BibitemShut {NoStop}%
\bibitem [{\citenamefont {Barouch}\ \emph {et~al.}(1970)\citenamefont
  {Barouch}, \citenamefont {McCoy},\ and\ \citenamefont
  {Dresden}}]{barouch1970}%
  \BibitemOpen
  \bibfield  {author} {\bibinfo {author} {\bibfnamefont {E.}~\bibnamefont
  {Barouch}}, \bibinfo {author} {\bibfnamefont {B.~M.}\ \bibnamefont {McCoy}},
  \ and\ \bibinfo {author} {\bibfnamefont {M.}~\bibnamefont {Dresden}},\ }\href
  {\doibase 10.1103/PhysRevA.2.1075} {\bibfield  {journal} {\bibinfo  {journal}
  {Phys. Rev. A}\ }\textbf {\bibinfo {volume} {2}},\ \bibinfo {pages} {1075}
  (\bibinfo {year} {1970})}\BibitemShut {NoStop}%
\bibitem [{\citenamefont {Barouch}\ and\ \citenamefont
  {McCoy}(1971{\natexlab{a}})}]{barouch1971}%
  \BibitemOpen
  \bibfield  {author} {\bibinfo {author} {\bibfnamefont {E.}~\bibnamefont
  {Barouch}}\ and\ \bibinfo {author} {\bibfnamefont {B.~M.}\ \bibnamefont
  {McCoy}},\ }\href {\doibase 10.1103/PhysRevA.3.786} {\bibfield  {journal}
  {\bibinfo  {journal} {Phys. Rev. A}\ }\textbf {\bibinfo {volume} {3}},\
  \bibinfo {pages} {786} (\bibinfo {year} {1971}{\natexlab{a}})}\BibitemShut
  {NoStop}%
\bibitem [{\citenamefont {Barouch}\ and\ \citenamefont
  {McCoy}(1971{\natexlab{b}})}]{barouch1971a}%
  \BibitemOpen
  \bibfield  {author} {\bibinfo {author} {\bibfnamefont {E.}~\bibnamefont
  {Barouch}}\ and\ \bibinfo {author} {\bibfnamefont {B.~M.}\ \bibnamefont
  {McCoy}},\ }\href {\doibase 10.1103/PhysRevA.3.2137} {\bibfield  {journal}
  {\bibinfo  {journal} {Phys. Rev. A}\ }\textbf {\bibinfo {volume} {3}},\
  \bibinfo {pages} {2137} (\bibinfo {year} {1971}{\natexlab{b}})}\BibitemShut
  {NoStop}%
\bibitem [{\citenamefont {Di~Giulio}\ \emph {et~al.}(2019)\citenamefont
  {Di~Giulio}, \citenamefont {Arias},\ and\ \citenamefont
  {Tonni}}]{digiulio2019}%
  \BibitemOpen
  \bibfield  {author} {\bibinfo {author} {\bibfnamefont {G.}~\bibnamefont
  {Di~Giulio}}, \bibinfo {author} {\bibfnamefont {R.}~\bibnamefont {Arias}}, \
  and\ \bibinfo {author} {\bibfnamefont {E.}~\bibnamefont {Tonni}},\
  }\href@noop {} {\  (\bibinfo {year} {2019})},\ \Eprint
  {http://arxiv.org/abs/1905.01144} {arXiv:1905.01144} \BibitemShut {NoStop}%
\bibitem [{\citenamefont {Calabrese}\ and\ \citenamefont
  {Cardy}(2006)}]{calabrese2006a}%
  \BibitemOpen
  \bibfield  {author} {\bibinfo {author} {\bibfnamefont {P.}~\bibnamefont
  {Calabrese}}\ and\ \bibinfo {author} {\bibfnamefont {J.}~\bibnamefont
  {Cardy}},\ }\href {\doibase 10.1103/PhysRevLett.96.136801} {\bibfield
  {journal} {\bibinfo  {journal} {Phys. Rev. Lett.}\ }\textbf {\bibinfo
  {volume} {96}},\ \bibinfo {pages} {136801} (\bibinfo {year}
  {2006})}\BibitemShut {NoStop}%
\bibitem [{\citenamefont {Calabrese}\ and\ \citenamefont
  {Cardy}(2007)}]{calabrese2007}%
  \BibitemOpen
  \bibfield  {author} {\bibinfo {author} {\bibfnamefont {P.}~\bibnamefont
  {Calabrese}}\ and\ \bibinfo {author} {\bibfnamefont {J.}~\bibnamefont
  {Cardy}},\ }\href {\doibase 10.1088/1742-5468/2007/06/P06008} {\bibfield
  {journal} {\bibinfo  {journal} {J. Stat. Mech.: Theory Exp.}\ }\textbf
  {\bibinfo {volume} {2007}},\ \bibinfo {pages} {P06008} (\bibinfo {year}
  {2007})}\BibitemShut {NoStop}%
\bibitem [{\citenamefont {Robinson}(1976)}]{robinson1976}%
  \BibitemOpen
  \bibfield  {author} {\bibinfo {author} {\bibfnamefont {D.~W.}\ \bibnamefont
  {Robinson}},\ }\href {\doibase 10.1017/S0334270000001260} {\bibfield
  {journal} {\bibinfo  {journal} {The Journal of the Australian Mathematical
  Society. Series B. Applied Mathematics}\ }\textbf {\bibinfo {volume} {19}},\
  \bibinfo {pages} {387} (\bibinfo {year} {1976})}\BibitemShut {NoStop}%
\bibitem [{\citenamefont {Bratteli}\ and\ \citenamefont
  {Robinson}(1981)}]{bratteli1981}%
  \BibitemOpen
  \bibfield  {author} {\bibinfo {author} {\bibfnamefont {O.}~\bibnamefont
  {Bratteli}}\ and\ \bibinfo {author} {\bibfnamefont {D.~W.}\ \bibnamefont
  {Robinson}},\ }\href@noop {} {{\selectlanguage {en}\emph {\bibinfo {title}
  {Operator {{Algebras}} and {{Quantum Statistical Mechanics II}}:
  {{Equilibrium States Models}} in {{Quantum Statistical Mechanics}}}}}},\
  Theoretical and {{Mathematical Physics}}\ (\bibinfo  {publisher}
  {{Springer-Verlag}},\ \bibinfo {address} {{Berlin Heidelberg}},\ \bibinfo
  {year} {1981})\BibitemShut {NoStop}%
\bibitem [{\citenamefont {Cardy}(2014)}]{cardy2014}%
  \BibitemOpen
  \bibfield  {author} {\bibinfo {author} {\bibfnamefont {J.}~\bibnamefont
  {Cardy}},\ }\href {\doibase 10.1103/PhysRevLett.112.220401} {\bibfield
  {journal} {\bibinfo  {journal} {Phys. Rev. Lett.}\ }\textbf {\bibinfo
  {volume} {112}},\ \bibinfo {pages} {220401} (\bibinfo {year}
  {2014})}\BibitemShut {NoStop}%
\bibitem [{\citenamefont {Cardy}(2016)}]{cardy2016g}%
  \BibitemOpen
  \bibfield  {author} {\bibinfo {author} {\bibfnamefont {J.}~\bibnamefont
  {Cardy}},\ }\href {\doibase 10.1088/1751-8113/49/41/415401} {\bibfield
  {journal} {\bibinfo  {journal} {J. Phys. A: Math. Theor.}\ }\textbf {\bibinfo
  {volume} {49}},\ \bibinfo {pages} {415401} (\bibinfo {year}
  {2016})}\BibitemShut {NoStop}%
\bibitem [{\citenamefont {Takayanagi}\ and\ \citenamefont
  {Ugajin}(2010)}]{takayanagi2010}%
  \BibitemOpen
  \bibfield  {author} {\bibinfo {author} {\bibfnamefont {T.}~\bibnamefont
  {Takayanagi}}\ and\ \bibinfo {author} {\bibfnamefont {T.}~\bibnamefont
  {Ugajin}},\ }\href {\doibase 10.1007/JHEP11(2010)054} {\bibfield  {journal}
  {\bibinfo  {journal} {J. High Energy Phys.}\ }\textbf {\bibinfo {volume}
  {2010}},\ \bibinfo {pages} {54} (\bibinfo {year} {2010})}\BibitemShut
  {NoStop}%
\bibitem [{\citenamefont {{da Silva}}\ \emph {et~al.}(2015)\citenamefont {{da
  Silva}}, \citenamefont {Lopez}, \citenamefont {Mas},\ and\ \citenamefont
  {Serantes}}]{dasilva2015a}%
  \BibitemOpen
  \bibfield  {author} {\bibinfo {author} {\bibfnamefont {E.}~\bibnamefont {{da
  Silva}}}, \bibinfo {author} {\bibfnamefont {E.}~\bibnamefont {Lopez}},
  \bibinfo {author} {\bibfnamefont {J.}~\bibnamefont {Mas}}, \ and\ \bibinfo
  {author} {\bibfnamefont {A.}~\bibnamefont {Serantes}},\ }\href {\doibase
  10.1007/JHEP04(2015)038} {\bibfield  {journal} {\bibinfo  {journal} {J. High
  Energy Phys.}\ }\textbf {\bibinfo {volume} {2015}},\ \bibinfo {pages} {38}
  (\bibinfo {year} {2015})}\BibitemShut {NoStop}%
\bibitem [{\citenamefont {Rigol}\ \emph {et~al.}(2006)\citenamefont {Rigol},
  \citenamefont {Muramatsu},\ and\ \citenamefont {Olshanii}}]{rigol2006}%
  \BibitemOpen
  \bibfield  {author} {\bibinfo {author} {\bibfnamefont {M.}~\bibnamefont
  {Rigol}}, \bibinfo {author} {\bibfnamefont {A.}~\bibnamefont {Muramatsu}}, \
  and\ \bibinfo {author} {\bibfnamefont {M.}~\bibnamefont {Olshanii}},\ }\href
  {\doibase 10.1103/PhysRevA.74.053616} {\bibfield  {journal} {\bibinfo
  {journal} {Phys. Rev. A}\ }\textbf {\bibinfo {volume} {74}},\ \bibinfo
  {pages} {053616} (\bibinfo {year} {2006})}\BibitemShut {NoStop}%
\bibitem [{\citenamefont {Rigol}\ \emph {et~al.}(2007)\citenamefont {Rigol},
  \citenamefont {Dunjko}, \citenamefont {Yurovsky},\ and\ \citenamefont
  {Olshanii}}]{rigol2007b}%
  \BibitemOpen
  \bibfield  {author} {\bibinfo {author} {\bibfnamefont {M.}~\bibnamefont
  {Rigol}}, \bibinfo {author} {\bibfnamefont {V.}~\bibnamefont {Dunjko}},
  \bibinfo {author} {\bibfnamefont {V.}~\bibnamefont {Yurovsky}}, \ and\
  \bibinfo {author} {\bibfnamefont {M.}~\bibnamefont {Olshanii}},\ }\href
  {\doibase 10.1103/PhysRevLett.98.050405} {\bibfield  {journal} {\bibinfo
  {journal} {Phys. Rev. Lett.}\ }\textbf {\bibinfo {volume} {98}},\ \bibinfo
  {pages} {050405} (\bibinfo {year} {2007})}\BibitemShut {NoStop}%
\bibitem [{\citenamefont {Vidmar}\ and\ \citenamefont
  {Rigol}(2016)}]{vidmar2016}%
  \BibitemOpen
  \bibfield  {author} {\bibinfo {author} {\bibfnamefont {L.}~\bibnamefont
  {Vidmar}}\ and\ \bibinfo {author} {\bibfnamefont {M.}~\bibnamefont {Rigol}},\
  }\href {\doibase 10.1088/1742-5468/2016/06/064007} {\bibfield  {journal}
  {\bibinfo  {journal} {J. Stat. Mech.: Theory Exp.}\ }\textbf {\bibinfo
  {volume} {2016}},\ \bibinfo {pages} {64007} (\bibinfo {year}
  {2016})}\BibitemShut {NoStop}%
\bibitem [{\citenamefont {{J. Surace}}\ \emph {et~al.}()\citenamefont {{J.
  Surace}}, \citenamefont {{L. Tagliacozzo}},\ and\ \citenamefont {{E.
  Tonni}}}]{j.suracea}%
  \BibitemOpen
  \bibfield  {author} {\bibinfo {author} {\bibnamefont {{J. Surace}}}, \bibinfo
  {author} {\bibnamefont {{L. Tagliacozzo}}}, \ and\ \bibinfo {author}
  {\bibnamefont {{E. Tonni}}},\ }\href@noop {} {\bibinfo  {journal} {in
  preparation}\ }\BibitemShut {NoStop}%
\bibitem [{\citenamefont {Cardy}(1986)}]{cardy1986a}%
  \BibitemOpen
\bibfield  {journal} {  }\bibfield  {author} {\bibinfo {author} {\bibfnamefont
  {J.}~\bibnamefont {Cardy}},\ }\href {\doibase 10.1016/0550-3213(86)90596-1}
  {\bibfield  {journal} {\bibinfo  {journal} {Nucl. Phys. B}\ }\textbf
  {\bibinfo {volume} {275}},\ \bibinfo {pages} {200} (\bibinfo {year}
  {1986})}\BibitemShut {NoStop}%
\bibitem [{\citenamefont {Cardy}(1989)}]{cardy1989}%
  \BibitemOpen
  \bibfield  {author} {\bibinfo {author} {\bibfnamefont {J.}~\bibnamefont
  {Cardy}},\ }\href {\doibase 10.1016/0550-3213(89)90521-X} {\bibfield
  {journal} {\bibinfo  {journal} {Nucl. Phys. B}\ }\textbf {\bibinfo {volume}
  {324}},\ \bibinfo {pages} {581} (\bibinfo {year} {1989})}\BibitemShut
  {NoStop}%
\bibitem [{\citenamefont {Calabrese}\ and\ \citenamefont
  {Cardy}(2016)}]{calabrese2016a}%
  \BibitemOpen
  \bibfield  {author} {\bibinfo {author} {\bibfnamefont {P.}~\bibnamefont
  {Calabrese}}\ and\ \bibinfo {author} {\bibfnamefont {J.}~\bibnamefont
  {Cardy}},\ }\href {\doibase 10.1088/1742-5468/2016/06/064003} {\bibfield
  {journal} {\bibinfo  {journal} {J. Stat. Mech.: Theory Exp.}\ }\textbf
  {\bibinfo {volume} {2016}},\ \bibinfo {pages} {064003} (\bibinfo {year}
  {2016})}\BibitemShut {NoStop}%
\bibitem [{\citenamefont {Peschel}(2004{\natexlab{a}})}]{peschel2004}%
  \BibitemOpen
  \bibfield  {author} {\bibinfo {author} {\bibfnamefont {I.}~\bibnamefont
  {Peschel}},\ }\href {\doibase 10.1088/1742-5468/2004/06/P06004} {\bibfield
  {journal} {\bibinfo  {journal} {J. Stat. Mech.: Theory Exp.}\ }\textbf
  {\bibinfo {volume} {2004}},\ \bibinfo {pages} {P06004} (\bibinfo {year}
  {2004}{\natexlab{a}})}\BibitemShut {NoStop}%
\bibitem [{\citenamefont {Ohmori}\ and\ \citenamefont
  {Tachikawa}(2015)}]{ohmori2015a}%
  \BibitemOpen
  \bibfield  {author} {\bibinfo {author} {\bibfnamefont {K.}~\bibnamefont
  {Ohmori}}\ and\ \bibinfo {author} {\bibfnamefont {Y.}~\bibnamefont
  {Tachikawa}},\ }\href {\doibase 10.1088/1742-5468/2015/04/P04010} {\bibfield
  {journal} {\bibinfo  {journal} {J. Stat. Mech.: Theory Exp.}\ }\textbf
  {\bibinfo {volume} {2015}},\ \bibinfo {pages} {P04010} (\bibinfo {year}
  {2015})}\BibitemShut {NoStop}%
\bibitem [{\citenamefont {Alba}\ \emph {et~al.}(2017)\citenamefont {Alba},
  \citenamefont {Calabrese},\ and\ \citenamefont {Tonni}}]{alba2017b}%
  \BibitemOpen
  \bibfield  {author} {\bibinfo {author} {\bibfnamefont {V.}~\bibnamefont
  {Alba}}, \bibinfo {author} {\bibfnamefont {P.}~\bibnamefont {Calabrese}}, \
  and\ \bibinfo {author} {\bibfnamefont {E.}~\bibnamefont {Tonni}},\ }\href
  {\doibase 10.1088/1751-8121/aa9365} {\bibfield  {journal} {\bibinfo
  {journal} {J. Phys. A: Math. Theor.}\ }\textbf {\bibinfo {volume} {51}},\
  \bibinfo {pages} {024001} (\bibinfo {year} {2017})}\BibitemShut {NoStop}%
\bibitem [{\citenamefont {Tonni}\ \emph {et~al.}(2018)\citenamefont {Tonni},
  \citenamefont {{Rodr{\'i}guez-Laguna}},\ and\ \citenamefont
  {Sierra}}]{tonni2018}%
  \BibitemOpen
  \bibfield  {author} {\bibinfo {author} {\bibfnamefont {E.}~\bibnamefont
  {Tonni}}, \bibinfo {author} {\bibfnamefont {J.}~\bibnamefont
  {{Rodr{\'i}guez-Laguna}}}, \ and\ \bibinfo {author} {\bibfnamefont
  {G.}~\bibnamefont {Sierra}},\ }\href {\doibase 10.1088/1742-5468/aab67d}
  {\bibfield  {journal} {\bibinfo  {journal} {J. Stat. Mech.: Theory Exp.}\
  }\textbf {\bibinfo {volume} {2018}},\ \bibinfo {pages} {043105} (\bibinfo
  {year} {2018})}\BibitemShut {NoStop}%
\bibitem [{\citenamefont {Wen}\ \emph {et~al.}(2018)\citenamefont {Wen},
  \citenamefont {Ryu},\ and\ \citenamefont {Ludwig}}]{wen2018a}%
  \BibitemOpen
  \bibfield  {author} {\bibinfo {author} {\bibfnamefont {X.}~\bibnamefont
  {Wen}}, \bibinfo {author} {\bibfnamefont {S.}~\bibnamefont {Ryu}}, \ and\
  \bibinfo {author} {\bibfnamefont {A.~W.~W.}\ \bibnamefont {Ludwig}},\ }\href
  {\doibase 10.1088/1742-5468/aae84e} {\bibfield  {journal} {\bibinfo
  {journal} {J. Stat. Mech.: Theory Exp.}\ }\textbf {\bibinfo {volume}
  {2018}},\ \bibinfo {pages} {113103} (\bibinfo {year} {2018})}\BibitemShut
  {NoStop}%
\bibitem [{\citenamefont {Evenbly}\ \emph {et~al.}(2010)\citenamefont
  {Evenbly}, \citenamefont {Pfeifer}, \citenamefont {Pic{\'o}}, \citenamefont
  {Iblisdir}, \citenamefont {Tagliacozzo}, \citenamefont {McCulloch},\ and\
  \citenamefont {Vidal}}]{evenbly2010b}%
  \BibitemOpen
  \bibfield  {author} {\bibinfo {author} {\bibfnamefont {G.}~\bibnamefont
  {Evenbly}}, \bibinfo {author} {\bibfnamefont {R.~N.~C.}\ \bibnamefont
  {Pfeifer}}, \bibinfo {author} {\bibfnamefont {V.}~\bibnamefont {Pic{\'o}}},
  \bibinfo {author} {\bibfnamefont {S.}~\bibnamefont {Iblisdir}}, \bibinfo
  {author} {\bibfnamefont {L.}~\bibnamefont {Tagliacozzo}}, \bibinfo {author}
  {\bibfnamefont {I.~P.}\ \bibnamefont {McCulloch}}, \ and\ \bibinfo {author}
  {\bibfnamefont {G.}~\bibnamefont {Vidal}},\ }\href {\doibase DOI:
  10.1103/PhysRevB.82.161107; eprintid: arXiv:0912.1642} {\bibfield  {journal}
  {\bibinfo  {journal} {Phys. Rev. B}\ }\textbf {\bibinfo {volume} {82}},\
  \bibinfo {pages} {161107} (\bibinfo {year} {2010})}\BibitemShut {NoStop}%
\bibitem [{\citenamefont {Evenbly}\ and\ \citenamefont
  {Vidal}(2014)}]{evenbly2014}%
  \BibitemOpen
  \bibfield  {author} {\bibinfo {author} {\bibfnamefont {G.}~\bibnamefont
  {Evenbly}}\ and\ \bibinfo {author} {\bibfnamefont {G.}~\bibnamefont
  {Vidal}},\ }\href {\doibase 10.1007/s10955-014-0983-1} {\bibfield  {journal}
  {\bibinfo  {journal} {J. Stat. Phys.}\ }\textbf {\bibinfo {volume} {157}},\
  \bibinfo {pages} {931} (\bibinfo {year} {2014})}\BibitemShut {NoStop}%
\bibitem [{\citenamefont {{A. Polkovnikov}}\ and\ \citenamefont {{M.
  Rigol}}()}]{a.polkovnikov}%
  \BibitemOpen
  \bibfield  {author} {\bibinfo {author} {\bibnamefont {{A. Polkovnikov}}}\
  and\ \bibinfo {author} {\bibnamefont {{M. Rigol}}},\ }\href@noop {} {\bibinfo
   {journal} {private communication}\ }\BibitemShut {NoStop}%
\bibitem [{\citenamefont {Torlai}\ \emph {et~al.}(2014)\citenamefont {Torlai},
  \citenamefont {Tagliacozzo},\ and\ \citenamefont {De~Chiara}}]{torlai2014}%
  \BibitemOpen
\bibfield  {journal} {  }\bibfield  {author} {\bibinfo {author} {\bibfnamefont
  {G.}~\bibnamefont {Torlai}}, \bibinfo {author} {\bibfnamefont
  {L.}~\bibnamefont {Tagliacozzo}}, \ and\ \bibinfo {author} {\bibfnamefont
  {G.}~\bibnamefont {De~Chiara}},\ }\href {\doibase
  10.1088/1742-5468/2014/06/P06001} {\bibfield  {journal} {\bibinfo  {journal}
  {J. Stat. Mech.: Theory Exp.}\ }\textbf {\bibinfo {volume} {2014}},\ \bibinfo
  {pages} {P06001} (\bibinfo {year} {2014})}\BibitemShut {NoStop}%
\bibitem [{\citenamefont {Bisognano}\ and\ \citenamefont
  {Wichmann}(1975)}]{bisognano1975}%
  \BibitemOpen
  \bibfield  {author} {\bibinfo {author} {\bibfnamefont {J.~J.}\ \bibnamefont
  {Bisognano}}\ and\ \bibinfo {author} {\bibfnamefont {E.~H.}\ \bibnamefont
  {Wichmann}},\ }\href {\doibase 10.1063/1.522605} {\bibfield  {journal}
  {\bibinfo  {journal} {J. Math. Phys.}\ }\textbf {\bibinfo {volume} {16}},\
  \bibinfo {pages} {985} (\bibinfo {year} {1975})}\BibitemShut {NoStop}%
\bibitem [{\citenamefont {Peschel}\ and\ \citenamefont
  {Chung}(1999)}]{peschel1999}%
  \BibitemOpen
  \bibfield  {author} {\bibinfo {author} {\bibfnamefont {I.}~\bibnamefont
  {Peschel}}\ and\ \bibinfo {author} {\bibfnamefont {M.-C.}\ \bibnamefont
  {Chung}},\ }\href {\doibase 10.1088/0305-4470/32/48/305} {\bibfield
  {journal} {\bibinfo  {journal} {J. Phys. A: Math. Gen.}\ }\textbf {\bibinfo
  {volume} {32}},\ \bibinfo {pages} {8419} (\bibinfo {year}
  {1999})}\BibitemShut {NoStop}%
\bibitem [{\citenamefont {Peschel}(2003)}]{peschel2003a}%
  \BibitemOpen
  \bibfield  {author} {\bibinfo {author} {\bibfnamefont {I.}~\bibnamefont
  {Peschel}},\ }\href {\doibase 10.1088/0305-4470/36/14/101} {\bibfield
  {journal} {\bibinfo  {journal} {J. Phys. A: Math. Gen.}\ }\textbf {\bibinfo
  {volume} {36}},\ \bibinfo {pages} {L205} (\bibinfo {year}
  {2003})}\BibitemShut {NoStop}%
\bibitem [{\citenamefont {Peschel}(2004{\natexlab{b}})}]{peschel2004a}%
  \BibitemOpen
  \bibfield  {author} {\bibinfo {author} {\bibfnamefont {I.}~\bibnamefont
  {Peschel}},\ }\href {\doibase 10.1088/1742-5468/2004/06/P06004} {\bibfield
  {journal} {\bibinfo  {journal} {J. Stat. Mech.: Theory Exp.}\ }\textbf
  {\bibinfo {volume} {2004}},\ \bibinfo {pages} {P06004} (\bibinfo {year}
  {2004}{\natexlab{b}})}\BibitemShut {NoStop}%
\bibitem [{\citenamefont {Peschel}\ and\ \citenamefont
  {Eisler}(2009)}]{peschel2009a}%
  \BibitemOpen
  \bibfield  {author} {\bibinfo {author} {\bibfnamefont {I.}~\bibnamefont
  {Peschel}}\ and\ \bibinfo {author} {\bibfnamefont {V.}~\bibnamefont
  {Eisler}},\ }\href {\doibase 10.1088/1751-8113/42/50/504003} {\bibfield
  {journal} {\bibinfo  {journal} {J. Phys. A: Math. Theor.}\ }\textbf {\bibinfo
  {volume} {42}},\ \bibinfo {pages} {504003} (\bibinfo {year}
  {2009})}\BibitemShut {NoStop}%
\bibitem [{\citenamefont {Casini}\ \emph {et~al.}(2011)\citenamefont {Casini},
  \citenamefont {Huerta},\ and\ \citenamefont {Myers}}]{casini2011a}%
  \BibitemOpen
  \bibfield  {author} {\bibinfo {author} {\bibfnamefont {H.}~\bibnamefont
  {Casini}}, \bibinfo {author} {\bibfnamefont {M.}~\bibnamefont {Huerta}}, \
  and\ \bibinfo {author} {\bibfnamefont {R.~C.}\ \bibnamefont {Myers}},\ }\href
  {\doibase 10.1007/JHEP05(2011)036} {\bibfield  {journal} {\bibinfo  {journal}
  {J. High Energy Phys.}\ }\textbf {\bibinfo {volume} {2011}},\ \bibinfo
  {pages} {36} (\bibinfo {year} {2011})}\BibitemShut {NoStop}%
\bibitem [{\citenamefont {Eisler}\ and\ \citenamefont
  {Peschel}(2017)}]{eisler2017a}%
  \BibitemOpen
  \bibfield  {author} {\bibinfo {author} {\bibfnamefont {V.}~\bibnamefont
  {Eisler}}\ and\ \bibinfo {author} {\bibfnamefont {I.}~\bibnamefont
  {Peschel}},\ }\href {\doibase 10.1088/1751-8121/aa76b5} {\bibfield  {journal}
  {\bibinfo  {journal} {J. Phys. A: Math. Theor.}\ }\textbf {\bibinfo {volume}
  {50}},\ \bibinfo {pages} {284003} (\bibinfo {year} {2017})}\BibitemShut
  {NoStop}%
\bibitem [{\citenamefont {Eisler}\ \emph {et~al.}(2019)\citenamefont {Eisler},
  \citenamefont {Tonni},\ and\ \citenamefont {Peschel}}]{eisler2019a}%
  \BibitemOpen
  \bibfield  {author} {\bibinfo {author} {\bibfnamefont {V.}~\bibnamefont
  {Eisler}}, \bibinfo {author} {\bibfnamefont {E.}~\bibnamefont {Tonni}}, \
  and\ \bibinfo {author} {\bibfnamefont {I.}~\bibnamefont {Peschel}},\ }\href
  {\doibase 10.1088/1742-5468/ab1f0e} {\bibfield  {journal} {\bibinfo
  {journal} {J. Stat. Mech.: Theory Exp.}\ }\textbf {\bibinfo {volume}
  {2019}},\ \bibinfo {pages} {073101} (\bibinfo {year} {2019})}\BibitemShut
  {NoStop}%
\bibitem [{\citenamefont {Fagotti}\ and\ \citenamefont
  {Essler}(2013)}]{fagotti2013b}%
  \BibitemOpen
  \bibfield  {author} {\bibinfo {author} {\bibfnamefont {M.}~\bibnamefont
  {Fagotti}}\ and\ \bibinfo {author} {\bibfnamefont {F.~H.~L.}\ \bibnamefont
  {Essler}},\ }\href {\doibase 10.1103/PhysRevB.87.245107} {\bibfield
  {journal} {\bibinfo  {journal} {Phys. Rev. B}\ }\textbf {\bibinfo {volume}
  {87}},\ \bibinfo {pages} {245107} (\bibinfo {year} {2013})}\BibitemShut
  {NoStop}%
\end{thebibliography}

%

\end{document}